%% file: main.tex
\documentclass[aps,prb,twocolumn,groupedaddress,superscriptaddress,noeprint]{revtex4-2}

\usepackage{orcidlink}
\usepackage{amsmath}
\usepackage{amssymb}
\usepackage{amsthm}
\usepackage{graphicx}
\usepackage{bm}
\usepackage{xcolor}
\usepackage{hyperref}
\usepackage{booktabs}
\usepackage{tikz}
\usetikzlibrary{positioning}

\newtheorem{theorem}{Theorem}
\newtheorem{proposition}[theorem]{Proposition}

\theoremstyle{remark}
\newtheorem*{remark}{Remark}

\newcommand{\lean}[1]{\texttt{#1}}
\newcommand{\Hop}{\hat H}
\newcommand{\R}{\mathbb{R}}

\usepackage[normalem]{ulem}
\makeatletter
\newcommand\raggedcells{\expandafter\def\expandafter\@arrayparboxrestore\expandafter{\@arrayparboxrestore\raggedright\let\\\tabularnewline}}
\makeatother
\newcommand\TableIcontinued{\end{tabular}\end{ruledtabular}\end{table*}\begin{table*}[!tp]\addtocounter{table}{-1}\caption{(Continued.)}\footnotesize\raggedcells\begin{ruledtabular}\begin{tabular}{p{4.2cm}p{5.0cm}p{7.6cm}}Assumption & Used in & Without it \\}
\newcommand\TableIIcontinued{\end{tabular}\end{ruledtabular}\end{table*}\begin{table*}[!tp]\addtocounter{table}{-1}\caption{(Continued.)}\footnotesize\raggedcells\begin{ruledtabular}\begin{tabular}{p{6.4cm}cp{9.2cm}}Result & Section & Lean name \\ \hline}

\begin{document}

\title{AI-assisted formal proofs for static and time-dependent density functional theory on lattices}

\newcommand{\bochum}{Research Center Future Energy Materials and Systems of the University Alliance Ruhr and Interdisciplinary Centre for Advanced Materials Simulation, Ruhr University Bochum, Universit\"atsstra{\ss}e 150, D-44801 Bochum, Germany}

\author{Miguel A. L. Marques\orcidlink{0000-0003-0170-8222}}
\email{miguel.marques@rub.de}
\affiliation{\bochum}
\author{Irene D'Amico\orcidlink{0000-0002-4794-1348}}
\affiliation{School of Physics, Engineering and Technology, University of York, YO10~5DD York, United Kingdom}
\affiliation{York Center for Quantum Technologies, University of York, YO10 5DD, York, United Kingdom}
\author{Stefano Pittalis\orcidlink{0000-0001-6941-5579}}
\affiliation{Istituto Nanoscienze -- CNR, S3, Via G. Campi 213/A, 41125 Modena, Italy}

\date{\today}

\begin{abstract}
The Hohenberg-Kohn theorem fails on lattices, where two potentials that differ by more than a constant can share a
ground state. We prove that such potentials are exceptional. We consider spinless fermions on any connected lattice, at any filling other than empty or full, with hopping amplitudes of a common
sign and any density-density interaction. The potentials whose ground state is shared with a potential
differing by more than a constant lie in a closed, nowhere dense set of zero Lebesgue measure. The densities of non-degenerate shared ground states form a set of codimension at least two. The statement about the potentials also holds for the Kohn-Sham potential and for spin-density functional theory, where a uniform shift of the
magnetic field is the only additional freedom. In time-dependent density functional theory on lattices, we show that not every ground state of a connected lattice can serve as the initial state of the density-potential map. A counterexample is the ground state of two fermions on a square that Penz and van~Leeuwen found as a counterexample to the Hohenberg-Kohn theorem. For almost every static potential, however, every real ground state can serve as an initial state, also for spin-resolved densities. We also give formal proofs that the time-dependent Kohn-Sham potential exists, locally in time, for almost every static Kohn-Sham potential, and of known results: local existence and uniqueness of the time-dependent potential for an admissible initial state, continuation in time up to the boundary of the set of admissible states, and the convergence properties of a regularized Kohn-Sham iteration. All theorems and propositions are formalized in the proof assistant Lean~4 and
checked by computer.
\end{abstract}

\maketitle

\section{Introduction}
\label{sec:intro}

It was in 1964 that Hohenberg and Kohn showed that the ground-state density of an interacting electron system
determines the external potential up to a constant~\cite{HohenbergKohn1964}. This theorem is the foundation of
density functional theory (DFT), together with the Kohn-Sham construction~\cite{KohnSham1965}. Its
time-dependent counterpart is the Runge-Gross theorem~\cite{RungeGross1984}, the foundation of time-dependent
DFT (TDDFT). The original proof of Hohenberg and Kohn uses one property of the continuum Schr\"odinger
equation that is easy to overlook. A ground state cannot vanish on a set of positive measure. On a lattice
this property is lost.

Lattice models are nevertheless the natural place to test the foundations of DFT. The Hilbert space is finite,
the density is a vector, and every functional can be computed exactly. Lattice DFT has a long
history~\cite{GunnarssonSchoenhammer1986,SchoenhammerGunnarssonNoack1995,ChayesChayesRuskai1985,CapelleCampo2013}.
However, Penz and van~Leeuwen showed in 2021 that on graphs the Hohenberg-Kohn theorem is ``found void, in
general''~\cite{PenzvanLeeuwen2021}. Two fermions on a square have a non-degenerate ground state, with all
site occupations strictly between zero and one, that is the ground state of a whole family of potentials. They
also proved that the densities produced by potentials that differ by more than a constant have measure zero [Theorem 11 of
Ref.~\onlinecite{PenzvanLeeuwen2021}], and stated that these densities form a closed set [Corollary 10 of Ref.~\onlinecite{PenzvanLeeuwen2023}; in Ref.~\onlinecite{PenzvanLeeuwen2021} it is given as what would follow from the conjecture that the uniquely $v$-representable densities form an open subset of the density domain], but this fails for two fermions on the square (Sec.~\ref{sec:densities}); the measure-zero result is unaffected.
The corresponding statement in potential space would say that the potentials affected by the failure are rare.
An early dimension-counting argument adapted from Ullrich and Kohn, given as an explorative outlook rather than a theorem [Sec.~IV.B of Ref.~\onlinecite{PenzvanLeeuwen2021}] led to what Penz and van~Leeuwen later called a ``premature conclusion'' [Sec.~V of Ref.~\onlinecite{PenzvanLeeuwen2023}]. To our
knowledge, the statement has not been proven for fermions, except for a single particle with $t>0$ [Corollary~15 of Ref.~\onlinecite{PenzvanLeeuwen2021}] or with uniform hopping [Proposition~3.6 of Ref.~\onlinecite{JamingMayeli2026}], and for the linear chain with $t>0$ [Theorem~17 of Ref.~\onlinecite{PenzvanLeeuwen2021}].

In TDDFT on lattices the situation is similar. Farzanehpour and Tokatly proved that a density with a continuous second time derivative is produced, locally in
time, by a unique potential~\cite{FarzanehpourTokatly2012}. This requires the initial state to belong to a set
$\Omega$, in which the matrix $K$ of the force-balance equation has no zero mode other than the constant potential. They also stated that every ground state of a connected lattice belongs to $\Omega$ (Sec.~V of
Ref.~\onlinecite{FarzanehpourTokatly2012}), and concluded in their Theorem 2 that, for a system starting in its
ground state, every density with a continuous second time derivative is locally $v$-representable. This statement was repeated in a later review~\cite{RuggenthalerPenzvanLeeuwen2015}. Recently, Canc\`es \textit{et al.} proved that $K$ is invertible at a ground state under two additional
hypotheses, non-degeneracy and a unique $v$-representability condition, which is the discrete analogue of unique
continuation [Theorem 5 and Eq.~(62) of Ref.~\onlinecite{CancesEtAl2026Lattice}]. They noted that the
latter ``is not always true'' in the discrete case and that the examples of Penz and van~Leeuwen suggest that it usually holds and is therefore natural to assume for ground states. As we show below, the counterexample of Penz and van~Leeuwen violates
this condition and contradicts Theorem 2 of Ref.~\onlinecite{FarzanehpourTokatly2012}. To our knowledge, the contradiction has not been noted before.

In this work we prove that both failures are exceptional. Both are tied to zero amplitudes in the configuration basis. We consider spinless fermions on any connected lattice, at any filling other than empty or full, with hopping amplitudes of a common sign
and any density-density interaction. For almost every potential, no eigenstate of the many-body Hamiltonian has a
zero amplitude.  We call a statement generic if it holds for every potential outside a closed set of zero measure. It then holds for almost every potential, and under small enough changes of any potential outside this set. In this work the closed set is the zero set of a nonzero polynomial in the potential, so a failure requires the potential to satisfy an exact equation. Such potentials need not be artificial: for two fermions on the square with $t=1$ and no interaction, the Hohenberg-Kohn theorem fails for every non-constant potential of the plane $v_1+v_3=v_2+v_4$; up to a constant, this plane is the two-parameter family found by Penz and van~Leeuwen (Sec.~\ref{sec:examples}). The key step is an algebraic statement about
the Krylov space of a single configuration, which we prove by taking the limit of a strong potential (as Jaming and Mayeli did for one particle~\cite{JamingMayeli2026}). There the
many-body sign problem reduces to the sign of a permutation, and a resonance argument selects a single
permutation. From this one result we obtain the generic Hohenberg-Kohn theorem for spinless fermions, for spin-DFT, and for
the Kohn-Sham potential. It also shows that the densities of non-degenerate ground states that are not uniquely v-representable have codimension at least two, which sharpens, for these densities, the measure-zero result of Penz and van~Leeuwen. We also obtain the corrected version of the theorem of Farzanehpour and Tokatly, which extends to spin-resolved densities. We further prove the existence of Kohn-Sham potentials, and local existence, uniqueness and continuation in
time of the time-dependent potential. We also prove a lattice version of van~Leeuwen's theorem~\cite{vanLeeuwen1999}, the local existence of the
time-dependent Kohn-Sham potential for pure and ensemble initial states\ (both for almost every static Kohn-Sham potential, where the non-interacting ground-state ensembles are pure), and the convergence properties of the
regularized Kohn-Sham iteration of Penz \textit{et al.}~\cite{Penz2019}. \endgraf Some of these results are known, and for them our contribution is the formal proof; the text gives the relation to the earlier statements. The existence of Kohn-Sham potentials is due to Refs.~\onlinecite{ChayesChayesRuskai1985,PenzvanLeeuwen2021}. Local existence and uniqueness of the time-dependent potential are part~(i) of Theorem~1 of Ref.~\onlinecite{FarzanehpourTokatly2012}. The lattice version of van~Leeuwen's theorem is this result, in the form of Theorem~\ref{thm:tdlocal}, applied to the second system with the density of the first; for a non-interacting second system it is noted in Ref.~\onlinecite{CancesDuez2026Dimer}, which attributes it to Refs.~\onlinecite{FarzanehpourTokatly2012,Tokatly2011}. Continuation in time is one direction of part~(ii) of Theorem~1 of Ref.~\onlinecite{FarzanehpourTokatly2012}. The convergence properties of the regularized Kohn-Sham iteration are those of the erratum~\cite{Penz2020Erratum} to Ref.~\onlinecite{Penz2019}. To our knowledge, what is new for the time-dependent Kohn-Sham potential is the generic statement of Theorem~\ref{thm:tdks}: for a pure initial state it exists, locally in time, for almost every static Kohn-Sham potential. The ensemble version applies the lattice van~Leeuwen theorem to a purified state.\ Relative to Penz and van~Leeuwen, who showed that the densities that are not uniquely $v$-representable have measure zero, we show that, under the assumptions above, the potentials responsible lie in a closed, nowhere dense set of zero measure; and for their square, where they found a two-parameter family of such potentials, we show that the exceptional set itself is not closed (the constant potentials are limits of exceptional potentials but are not exceptional) and give a necessary condition on the hopping. Several of the results listed here hold beyond the setting stated above; they are collected at the end of this section.

The proofs of all theorems and propositions in this work are written in the proof assistant Lean~4~\cite{Lean4}, with the mathematical library
Mathlib~\cite{Mathlib}. Proof assistants are a relatively new technology. A proof is accepted only if every step
follows from the axioms of mathematics, and the check is done by computer. This puts the correctness of the
proofs beyond doubt. What the computer cannot check is whether the formal statements
say what we mean. We discuss this question, and what is not proved, in Sec.~\ref{sec:formal}. The code is
openly available~\cite{leandft}. To our knowledge, this is the first formalization of DFT in a proof assistant,
although Lean has been used before for other topics in physics and chemistry~\cite{Bobbin2024,ToobySmith2025HepLean}.

The paper is organized as follows. In Sec.~\ref{sec:setting} we define lattice DFT for spinless fermions and explain why the
Hohenberg-Kohn theorem fails. Section~\ref{sec:generic} contains the generic Hohenberg-Kohn theorem, an outline of its proof, and the dimension bound for the non-uniquely v-representable densities. Section~\ref{sec:spin} extends the theorem to spin-DFT and treats potentials that couple to the total density of spinful electrons, for which the Hohenberg-Kohn theorem holds for every potential when there is one electron of each spin and the hopping is uniform with $t>0$. Section~\ref{sec:ks} contains the static Kohn-Sham results. Time-dependent DFT is the subject of Sec.~\ref{sec:td}, which also contains a theorem on static ground states (Theorem~\ref{thm:gsomega}). It treats the counterexample to Theorem~2 of Ref.~\onlinecite{FarzanehpourTokatly2012} (Sec.~\ref{sec:counter}), spinful electrons (Sec.~\ref{sec:spintd}), and van~Leeuwen's theorem and the time-dependent Kohn-Sham potential (Sec.~\ref{sec:tdks}). In Sec.~\ref{sec:assumptions} we list all assumptions
and the counterexamples that appear when they are dropped. In Sec.~\ref{sec:formal} we describe the
formalization, and in Sec.~\ref{sec:conclusions} we draw our conclusions. Table~\ref{tab:results} lists the
Lean name of every result. \endgraf The main results that hold beyond the setting stated above are the following. Local existence, uniqueness and continuation (Theorems~\ref{thm:tdlocal} and~\ref{thm:global}) hold for any real symmetric Hamiltonian with a fixed particle number and an initial state in $\Omega$. The lattice version of van~Leeuwen's theorem (Sec.~\ref{sec:tdks}) holds for any two real symmetric Hamiltonians with a fixed particle number on the same sites. The first system is driven by a continuous potential, and the second starts from a state in its own $\Omega$ with the same initial density and current. The existence of the Kohn-Sham potential (Theorem~\ref{thm:ksexist}) holds on any graph, connected or not, and its proof works for any real symmetric $\hat H_0$ with a fixed particle number. The regularized Kohn-Sham iteration (Theorem~\ref{thm:myksoda}) holds for any real symmetric interacting and reference Hamiltonians on the same configurations. The conclusions of Theorems~\ref{thm:ghk} and~\ref{thm:dens} also hold for some hopping that is not of a common sign: under a condition on the signs of the shortest paths, and on the square for every nonzero hopping (Remark in Sec.~\ref{sec:generic}).

\section{Lattice DFT and the failure of the Hohenberg-Kohn theorem}
\label{sec:setting}

\subsection{Setting}

We consider $N$ spinless fermions on a lattice, i.e., a finite connected graph $G$ with sites $i=1,\dots,M$
and $0<N<M$. Spinful electrons are treated in Secs.~\ref{sec:spin} and~\ref{sec:spintd}. The Hamiltonian is
\begin{equation}
  \Hop(v) = \hat T + \hat W + \sum_{i=1}^M v_i \hat n_i ,
  \qquad
  \hat T = -\sum_{\langle ij\rangle} t_{ij}\left(\hat c_i^\dagger \hat c_j + \hat c_j^\dagger \hat c_i\right),
  \label{eq:ham}
\end{equation}
where the sum runs over the edges of $G$ and $t_{ij}=t_{ji}$ are real hopping amplitudes. For Theorems~\ref{thm:ghk}--\ref{thm:spin}, \ref{thm:ksunique}, \ref{thm:gsomega} and~\ref{thm:tdks} we assume that all
$t_{ij}$ have the same sign\ (for the conclusions of Theorems~\ref{thm:ghk} and~\ref{thm:dens} a weaker condition on the signs suffices). This assumption is used only in the proof of the key algebraic statement of
Sec.~\ref{sec:generic} (see the Remark at the end of the outline of the proof in that section). The main assumptions of this work, and what
happens when they are dropped, are collected in Sec.~\ref{sec:assumptions}. The interaction $\hat W$ is any operator diagonal in the occupation-number basis,
for example a nearest-neighbor or a long-range density-density interaction. The basis states $|c\rangle$ are
labeled by configurations $c$, i.e., sets of $N$ occupied sites, and $n_i(c)\in\{0,1\}$ is the occupation of
site $i$ in $c$. In this basis the potential term is diagonal: its entry for the configuration $c$ is $\sum_i v_in_i(c)=\sum_{i\in c}v_i$, the sum of the potential over the occupied sites (e.g.\ $v_1+v_4$ for $c=\{1,4\}$); only for one particle does it reduce to a single $v_i$. The Hamiltonian is a real symmetric matrix of dimension $\binom{M}{N}$, and we take
wavefunctions $\Psi(c)$ real. Complex states appear in the time-dependent theory of Sec.~\ref{sec:td}.

A ground state of $\Hop(v)$ is any nonzero minimizer of the Rayleigh quotient
$\langle\Psi|\Hop(v)|\Psi\rangle/\langle\Psi|\Psi\rangle$. We do not assume that the ground state is
non-degenerate, except in the bound on the non-uniquely v-representable densities (Theorem~\ref{thm:dens}, Sec.~\ref{sec:densities}) and in the segment $|\varepsilon|\le g/2$ of potentials that share the ground state of the square in Sec.~\ref{sec:examples}, where a gap is assumed. The density of a normalized state is $\rho_i=\sum_c n_i(c)\,\Psi(c)^2$, and
$\sum_i \rho_i = N$. The density thus consists of site occupations: on a real-space grid they approximate the number of particles in a grid cell, i.e., the continuum density times the cell volume, while in tight-binding models such as the Hubbard model they are occupations of localized orbitals. Adding a constant to $v$ changes neither the ground states nor the density.

Our formal statements are written for the matrix $\Hop(v)$ in the configuration basis. The fermionic signs
enter through the second quantization of the single-particle Hamiltonian, which we define with Slater
determinants (Sec.~\ref{sec:formal}).

\subsection{Weak and strong Hohenberg-Kohn theorems}

The first half of the argument of Hohenberg and Kohn holds on every lattice. Let $\Hop(v)$ and $\Hop(v')$ have normalized ground states $\Psi$ and $\Psi'$ with the same density. Then $\Psi$
is also a ground state of $\Hop(v')$, and $\Psi'$ is a ground state of $\Hop(v)$ (Theorem 7 of
Ref.~\onlinecite{PenzvanLeeuwen2021}). We call this result the weak Hohenberg-Kohn theorem; similar names are used in the literature (see Ref.~\onlinecite{PenzvanLeeuwen2023} and references therein).
Hence all potentials that have a ground state with a given density have the same ground states with this density. For degenerate ground states in the continuum, Capelle, Ullrich and Vignale gave the same variational argument (their joint-degeneracy theorem)~\cite{CapelleUllrichVignale2007}. The question is whether a shared ground state
determines the potential. The strong Hohenberg-Kohn theorem is the statement that two potentials with a common ground state differ only by a constant.

Let $\Psi$ be a ground state of both $\Hop(v)$ and $\Hop(v')$, with energies $E$ and $E'$. Subtracting the
two eigenvalue equations gives, for every configuration $c$,
\begin{equation}
  \Big[\sum_i (v_i-v_i')\,n_i(c) - (E-E')\Big]\Psi(c) = 0 .
  \label{eq:support}
\end{equation}
The function $w=v-v'$ is therefore ``invisible'', in the sense that $\sum_i w_i n_i(c)$ is constant on the
configurations where $\Psi$ does not vanish. If $\Psi$ has no zero amplitude, this holds for every configuration. For $0<N<M$
this forces $w$ to be constant. To see this, take two sites $i\neq j$ and a configuration that contains $i$ but
not $j$. Moving the particle from $i$ to $j$ changes $\sum_k w_k n_k(c)$ by $w_j-w_i$. We conclude that the
Hohenberg-Kohn theorem can fail only for ground states with zero amplitudes. This is the lattice counterpart of
the unique-continuation property of the continuum.

A similar argument was given for fermionic lattice models in Ref.~\cite{CoeDAmicoFranca2015}, and for graphs in Ref.~\onlinecite{PenzvanLeeuwen2021} [Eqs.~(35)--(37)].\footnote{Penz and van~Leeuwen also gave examples of ground states with vanishing components for which the Hohenberg-Kohn theorem fails, and pointed out that the non-vanishing of the components, on which such an argument relies, is not guaranteed in general.}
The proof of Ref.~\cite{CoeDAmicoFranca2015} is designed for wave functions that contain, with non-zero coefficient, ``no matter how small'', pairs of terms that differ only by the occupation of two sites, such as $|{\uparrow\downarrow},S,X\rangle$ and $|S,{\uparrow\downarrow},X\rangle$, or $|{\uparrow\downarrow},0,X\rangle$ and $|0,{\uparrow\downarrow},X\rangle$, or $|S,0,X\rangle$ and $|0,S,X\rangle$, where the first two entries are the two sites to compare, $S$ denotes a singly occupied site and $0$ an empty one, with the same configuration $X$ of the remaining particles. For spinless fermions only the third kind of pair occurs. When such pairs are present for site pairs that connect all the sites, the argument shows that the potential is determined up to a constant. The square example of Sec.~\ref{sec:examples}, with two spinless fermions, lies outside this class: such pairs exist only for the site pairs 1--2 and 3--4, which do not connect all the sites. Yet no site has zero density or is full, so the density does not reveal the zero amplitudes. The proof of Theorem~\ref{thm:ghk} completes the argument for spinless fermions with hopping of a common sign: for almost every potential the ground state has no zero amplitude, so the required terms are present and, as concluded in Ref.~\cite{CoeDAmicoFranca2015}, the density determines the potential up to a constant.
For electrons with spin and a spin-independent potential, as in the Hubbard model, we have shown the absence of zero amplitudes only for one electron of each spin, with uniform hopping $t>0$ on a connected graph, and there for every potential (Sec.~\ref{sec:spin}). This covers the square of Sec.~\ref{sec:examples} in the sector with one electron of each spin, which includes the two-electron ground state of the Hubbard model; there the density determines the potential up to a constant, in agreement with Ref.~\cite{CoeDAmicoFranca2015}.

\subsection{Examples}
\label{sec:examples}

For a single particle with $t>0$ the ground state has no zeros [Corollary~15 of Ref.~\onlinecite{PenzvanLeeuwen2021}], and the Hohenberg-Kohn theorem holds for every
potential. Our proof avoids the Perron-Frobenius theorem. If $\psi$ is a ground state, so is $|\psi|$, and at a
zero of $|\psi|$ the eigenvalue equation forces all neighbors to vanish, so that $|\psi|\equiv0$ by
connectivity. With frustrated hopping the theorem fails already for one particle. On the triangle with $t=-1$
there are two potentials, differing by more than a constant, with the same non-degenerate ground state. A one-particle example on the kagome lattice with $t=-1$, a compact localized state that remains the ground state for a whole family of potentials, is given in Ref.~\onlinecite{Xu2022}; these potentials vanish on a hexagon and form a set of measure zero. For random potentials on this lattice, Ref.~\onlinecite{Xu2022} reports a numerically vanishing smallest eigenvalue of its criterion matrix and reads it as a generic failure of the Hohenberg-Kohn theorem; by Theorem~\ref{thm:ghk} with $N=1$ (uniform \mbox{$t=-1$} is of a common sign), the Hohenberg-Kohn theorem holds for almost every potential on this lattice.

For two fermions on the triangle with $t=1$ and no interaction, the ground state $(1,0,-1)$ in the basis $\{12,13,23\}$ of potential
$v=(0,-\tau,0)$, $\tau>0$, never places the two particles on sites 1 and 3. Site 2 is always occupied, and
its potential is invisible [Eqs.~(42) and (43) of Ref.~\onlinecite{PenzvanLeeuwen2021}]. This failure is trivial, since the density of site 2 is one.

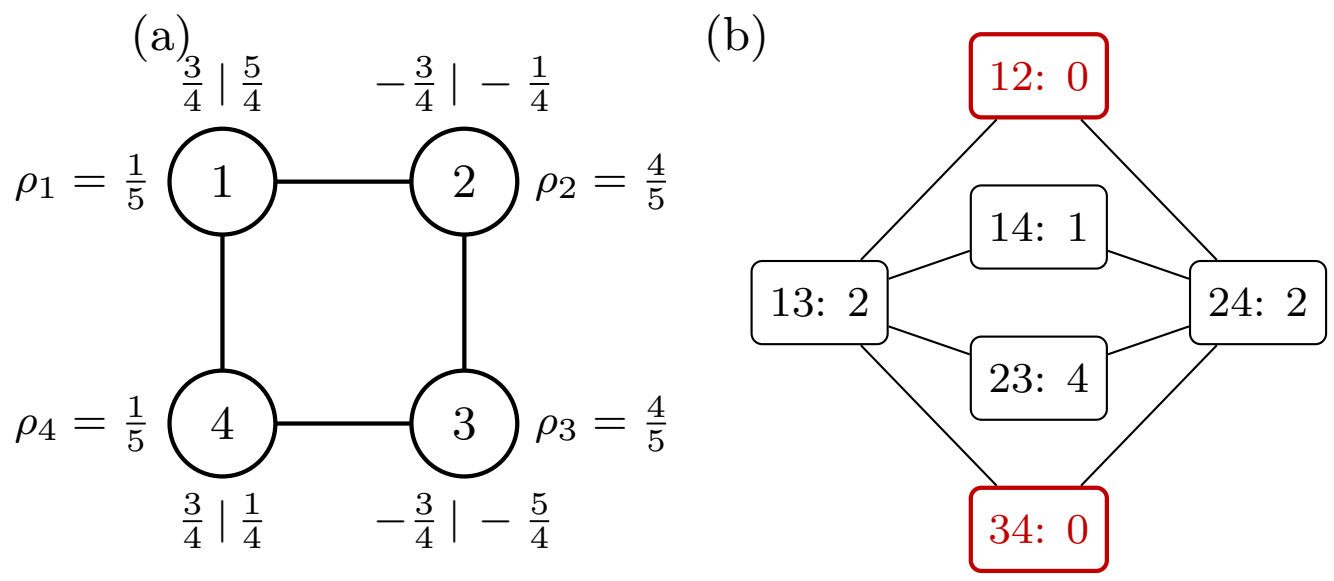
\begin{figure}[t]
  \centering
  \input{figures/square.tex}
  \caption{Two spinless fermions on the square (hopping $t=1$). (a) The two potentials $v$ and $v'$ (left and
  right of the bar) and the density of their common ground state. (b) The six configurations, connected by a
  single hop, with the ground-state amplitudes in units of $1/5$. The amplitudes on $\{1,2\}$ and
  $\{3,4\}$ vanish (red).}
  \label{fig:square}
\end{figure}

The counterexample of Penz and van~Leeuwen on the square is not trivial [state $\Psi_A$ of Eq.~(48) in
Ref.~\onlinecite{PenzvanLeeuwen2021}, with $s=3/4$]. Two fermions on the square
with $t=1$ and no interaction, in the potentials $v=(\tfrac34,-\tfrac34,-\tfrac34,\tfrac34)$ and
$v'=(\tfrac54,-\tfrac14,-\tfrac54,\tfrac14)$, have the same non-degenerate ground state, with energy $-5/2$ ($3/2$ in the convention of Ref.~\onlinecite{PenzvanLeeuwen2021}, whose Hamiltonian contains the constant $4$)
and gaps $1/2$ and $0.264$ to the first excited state. Its amplitudes are shown in Fig.~\ref{fig:square}. The
density is $(1,4,4,1)/5$, so no site is empty or full. The ground state vanishes on the configurations
$\{1,2\}$ and $\{3,4\}$, and $w=v'-v\propto(1,1,-1,-1)$ gives $\sum_i w_i n_i(c)=0$ on the other four
configurations. In fact, the failure covers the plane $v_1+v_3=v_2+v_4$ except for its constant potentials; up to a constant, this plane is the two-parameter family of potentials of Eq.~(45) of Ref.~\onlinecite{PenzvanLeeuwen2021} [their Secs.~III~C and IV~B, Fig.~4]. For every potential with $v_1+v_3=v_2+v_4$ and a
ground state with a gap $g>0$, the same ground state is shared at least by the segment $v+\varepsilon w$,
$|\varepsilon|\le g/2$, with $w=(1,1,-1,-1)$ if $|v_1-v_2|>|v_1-v_4|$, as above, and $w=(1,-1,-1,1)$ if $|v_1-v_2|<|v_1-v_4|$. We verified the ground-state property of both potentials with exact
sum-of-squares certificates.\footnote{A certificate is explicit data that makes a claim easy to check. Here, for each potential, $\langle\chi|\Hop(v)-E_0|\chi\rangle$, with $E_0$ the energy of the state, is written for every state $\chi$ as a sum, with positive rational weights, of squares of linear combinations of the amplitudes of $\chi$ with exact rational coefficients. A sum of squares cannot be negative, so no state has an energy below $E_0$. Lean checks this identity. The term is common in optimization and in computer-assisted proofs.}\ For the example above, where $g/2=1/4$, $\Psi_A$ is a ground state of $v+\varepsilon(1,1,-1,-1)$ exactly for $|\varepsilon|\le3/4$, three times $g/2$. In general ($t=1$, $\hat W=0$), on the plane the state $\Psi_A$ of Eq.~(48) of Ref.~\onlinecite{PenzvanLeeuwen2021}, with $s=(v_1-v_2)/2$, is a ground state of $v+\varepsilon(1,1,-1,-1)$ if and only if $|v_1-v_4+2\varepsilon|\le|v_1-v_2|$; outside this range it is still an eigenstate of $v+\varepsilon(1,1,-1,-1)$, but not a ground state (\lean{square\_example\_segment}, \lean{square\_segment\_plus}, \lean{eigen\_sq\_plus}). For electrons with spin, $\Psi_A$ is a state of two electrons with parallel spins (a triplet state); in the two-electron Hubbard model with the same potential it is an excited state for every $U$, the ground state being a singlet.\endgraf The zeros of $\Psi_A$ are tied to a symmetry of the hopping. For the Hamiltonian~(\ref{eq:ham}) on the square with arbitrary real $t_{ij}=t_{ji}$ on the four bonds, any potential and any interaction $\hat W$, an eigenstate that vanishes on $\{1,2\}$ and $\{3,4\}$, with $\Psi(\{1,3\})$ and $\Psi(\{2,4\})$ not both zero, exists only if $|t_{23}|=|t_{41}|$ (one that vanishes on $\{1,4\}$ and $\{2,3\}$ instead, again with $\Psi(\{1,3\})$ and $\Psi(\{2,4\})$ not both zero, only if $|t_{12}|=|t_{34}|$); and a ground state shared by two potentials $v$ and $v+\varepsilon(1,1,-1,-1)$, $\varepsilon\neq0$, exists only if $|t_{23}|=|t_{41}|$ or $t_{12}=t_{34}=0$, so with four nonzero hoppings only if $|t_{23}|=|t_{41}|$ (\lean{Bridge.Square.square\_zero\_pattern\_hopping}, \lean{Bridge.Square.square\_zero\_pattern\_hopping\textquotesingle}, \lean{Bridge.Square.square\_shared\_hopping\_sharp}). We do not show that these conditions are sufficient.

\section{Generic Hohenberg-Kohn theorem}
\label{sec:generic}

\subsection{Statement}

We call a potential $v$ \emph{exceptional} if some ground state of $\Hop(v)$ is also a ground state of
$\Hop(v')$ for a potential $v'$ that differs from $v$ by more than a constant.

\begin{theorem}[Generic Hohenberg-Kohn theorem]
\label{thm:ghk}
Let $G$ be a connected graph with $M$ sites, $0<N<M$, $t_{ij}=t_{ji}$ with $s\,t_{ij}>0$ on every edge for a
fixed sign $s$, and $\hat W$ any configuration-diagonal interaction. Then the exceptional potentials lie in a
closed subset of $\R^M$ with empty interior and zero Lebesgue measure.
\end{theorem}

Combined with the weak theorem of the previous section, this means that for every potential outside this set
the ground-state density determines the potential up to a constant. This set is closed and nowhere dense, so the density determines the potential on an open, dense set of potentials. The exceptional set is contained in this closed set and therefore has measure zero, so a potential chosen at random is
non-exceptional with probability one \footnote{In plain terms, the theorem does not say that the exceptional potentials form a closed set. It says that they lie inside a closed set, which can be larger. ``Closed'' means that the set contains all its limit points: if a sequence of potentials in the set converges, the limit is in the set. Because the set is closed, ``nowhere dense'' means that it contains no ball of potentials, however small. So an arbitrarily small change of an exceptional potential can make it non-exceptional, and a potential outside the closed set stays non-exceptional under all sufficiently small changes. ``Zero Lebesgue measure'' means that a potential whose $M$ site values are drawn at random from any distribution with a probability density lies outside the closed set, and is therefore non-exceptional, with probability one. The same holds when the potential is fixed only up to a constant, for example with $\sum_iv_i=0$, because adding a constant to the potential does not change the closed set (this follows from the proof). The exceptional set itself need not be closed: for two fermions on the square with $t=1$ and $\hat W=0$, the constant potentials are not exceptional, but they are limits of exceptional ones (end of the paragraph after Theorem~\ref{thm:ghk}).}.
The theorem covers degenerate ground states, and it does not require the interaction to be small. For two fermions on the square, the exceptional set contains every non-constant potential of the plane $v_1+v_3=v_2+v_4$ discussed above, so the theorem cannot be
improved to a statement about every potential.\ In this example ($t=1$, $\hat W=0$) the constant potentials are not exceptional, although they are limits of exceptional ones, so the exceptional set itself is not closed.

\subsection{Outline of the proof}

\begin{figure}[t]
  \centering
  \input{figures/proof.tex}
  \caption{Structure of the proof of the generic Hohenberg-Kohn theorem. The left column reduces the theorem
  to the hypothesis (H$^*$), and the right column proves (H$^*$) in the strong-potential limit.}
  \label{fig:proof}
\end{figure}
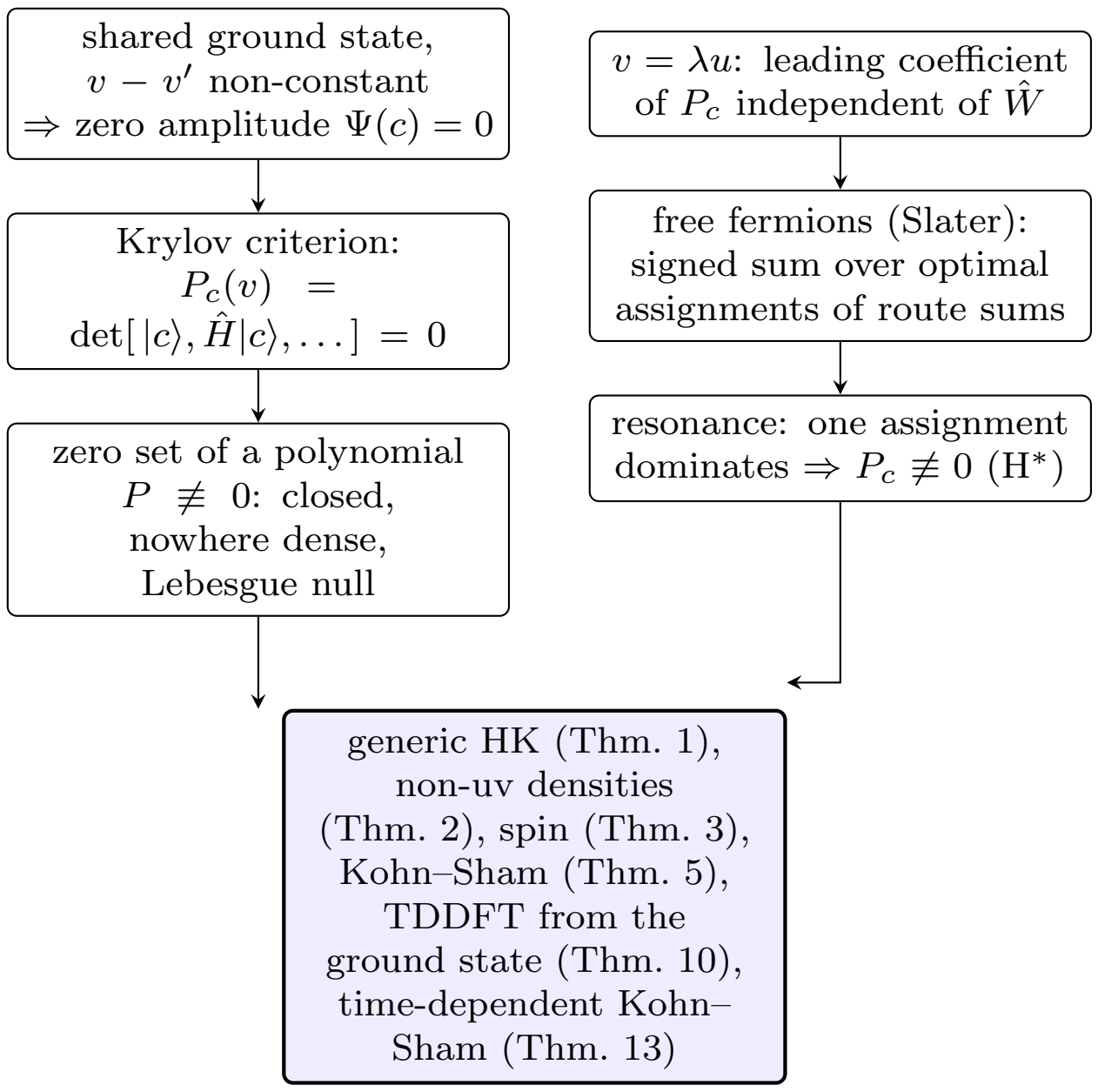

The structure of the proof is shown in Fig.~\ref{fig:proof}.

\emph{Krylov criterion.} Let $|c\rangle$ be a configuration and consider the $L\times L$ matrix
$K_c(v)=\big[\,|c\rangle,\ \Hop(v)|c\rangle,\ \dots,\ \Hop(v)^{L-1}|c\rangle\,\big]$, with $L=\binom MN$. If an
eigenstate $\Psi$ of $\Hop(v)$ satisfies $\Psi(c)=0$, then $\Psi$ is orthogonal to every column of $K_c(v)$,
since $\langle\Psi|\Hop^k|c\rangle=E^k\Psi(c)=0$. Hence $\det K_c(v)=0$. The determinant
$P_c(v)=\det K_c(v)$ is a polynomial in the potential. Physically, $P_c\neq0$ means that the dynamics started
from configuration $c$ reaches every eigenstate. For one particle with uniform hopping of either sign on a connected graph (the sign can be reversed by replacing $\Hop$ with $-\Hop$ and $v$ with $-v$), this criterion, together with a strong-potential limit as below, shows that for almost every potential every eigenvector has full support [Proposition~3.6 of Ref.~\onlinecite{JamingMayeli2026} and references therein]. For $N$ fermions the leading coefficient of this limit is a signed sum over permutations, and the resonance argument below is needed to show that it does not vanish.

\emph{Reduction.} By Eq.~\eqref{eq:support}, an exceptional potential has a ground state with a zero
amplitude, so it is a zero of $\prod_c P_c$. The zero set of a polynomial that does not vanish identically is
closed and nowhere dense, and it has zero Lebesgue measure (by induction on the number of variables and
Fubini's theorem). The theorem therefore follows from the hypothesis
\begin{equation}
  \text{(H$^*$)}\qquad P_c\not\equiv 0 \quad\text{for every configuration } c .
\end{equation}
Note that (H$^*$) also has a consequence for excited states. Outside the zero set the spectrum is simple, and no
eigenstate has a zero amplitude. A related argument was proposed by Xu \textit{et al.}, with the determinant of their generalized density correlation matrix, evaluated in the ground state, which they argue is typically analytic in the parameters: if it is nonzero at one point, a generalized Hohenberg-Kohn theorem is expected to hold generically~\cite{Xu2022}. The Krylov determinant $P_c$ is a polynomial in $v$ and needs no analyticity of the ground state.

\emph{Strong-potential limit.} We restrict the potential to a line $v=\lambda u$ and study the leading
power of $\lambda$ in $P_c(\lambda u)$. Let $d(y)$ be the minimal number of hops needed to move the particles of configuration $c$ to the sites of
configuration $y$. This is the cost of an optimal assignment of the particles, with the graph distance as
cost. A single hop changes $d$ by at most one, so that the matrix element
$\langle y|\Hop(\lambda u)^k|c\rangle$ is a polynomial in $\lambda$ of degree at most $k-d(y)$. The
coefficient of the highest power of $\lambda$ in $P_c(\lambda u)$ is therefore a sum over sequences of hops
that never waste a step. In such a sequence every diagonal factor contributes $\lambda$, so the interaction
$\hat W$, which carries no factor $\lambda$, drops out of the leading coefficient. It is enough to show that
the leading coefficient is nonzero for free fermions.

\emph{Free fermions.} Without interaction the leading coefficient factorizes into a Vandermonde determinant
of the configuration energies $\sum_{i\in y}u_i$, which is nonzero for generic $u$, and a product over
configurations $y$ of coefficients $\alpha_y(u)$. Using Slater determinants we show that
\begin{equation}
  \alpha_y(u) = \sum_{\pi\in\mathrm{Opt}(y,c)} \operatorname{sgn}(\pi)\prod_{l=1}^N
  \beta_{y_l\to c_{\pi(l)}}(u) ,
  \label{eq:alpha}
\end{equation}
where the sum runs over the optimal assignments between the sites of $y$ and those of $c$. The factor
$\beta_{a\to b}(u)$ is a single-particle route sum. It is a sum over the shortest paths from $a$ to $b$ of the
product of the hopping amplitudes, divided by the product of the differences $u_a-u_z$ over the sites $z\neq a$
of the path. Equation~\eqref{eq:alpha} is the many-body sign problem in its simplest form. The terms
carry the sign of the permutation, and they could cancel.

\emph{Resonance.} To show that they do not, we choose an optimal assignment $\pi_0$ that leaves the sites common to $c$ and $y$
in place. We then choose a potential $u^*$ in which every moved particle is degenerate with its target under
$\pi_0$. Along $u=u^*+\varepsilon\delta$, the term of $\pi_0$ has a pole of order equal to the
number of moved particles, and every other optimal assignment has a pole of lower order. The coefficient of
the leading pole is a product of route sums. Each of them is nonzero. When the corresponding pair of sites is moved to a large potential, all shortest paths
contribute with the same sign. This is where the sign condition on the hopping enters. Hence $\alpha_y\not\equiv0$. Finally, a finite product of nonzero
rational functions is nonzero somewhere, so a single $u$ makes all $\alpha_y$ and the Vandermonde determinant
nonzero, and (H$^*$) follows.

\begin{remark}
The sign condition is used only to show that the route sums do not vanish. Uniform hopping $t\neq0$ of either
sign satisfies it. Frustrated hopping is not excluded by the counterexample on the triangle. That example shows
that the theorem fails for some potentials, which is compatible with Theorem~\ref{thm:ghk} for $t=-1$. The argument needs less: it is enough that the hopping is nonzero and that for any two sites all shortest paths between them have hopping products of the same sign, i.e., that the signs of the $t_{ij}$ form a distance-compatible signed graph~\cite{HameedEtAl2021}. In the form that we formalized: for every site $b$ there are signs $\varepsilon_q=\pm1$ with $\varepsilon_qt_{qr}\varepsilon_r>0$ on every edge $qr$ along which $r$ is one step closer to $b$ than $q$; this forces $t_{qr}\neq0$ and is then equivalent to the form with paths (not formalized). Under this condition, (H$^*$) holds, and with it the conclusions of Theorems~\ref{thm:ghk} and~\ref{thm:dens} (\lean{hstar\_lattice\_fermions\_shortest}, \lean{generic\_hk\_lattice\_fermions\_shortest}, \lean{dimH\_nonUVDens\_lattice\_fermions\_shortest\textquotesingle}). Theorem~\ref{thm:ksunique} is the case $\hat W=0$ of Theorem~\ref{thm:ghk}, so it also holds under this condition (\lean{generic\_hk\_lattice\_fermions\_shortest}{} with $\hat W=0$); we have not restated the other generic theorems. The condition holds for every hopping that sign changes of the orbitals bring to a common sign (\lean{shortestPathSigned\_of\_gauge}; balanced and antibalanced signed graphs in Ref.~\onlinecite{HameedEtAl2021}). Such a change, $\hat c_i\to\sigma_i\hat c_i$ with $\sigma_i=\pm1$ (a gauge transformation with phases $0$ or $\pi$), multiplies $t_{ij}$ by $\sigma_i\sigma_j$ and leaves the densities, the potential, the interaction and the exceptional potentials unchanged (\lean{gamma\_signConj}, \lean{H0Fw\_gaugeHop}, \lean{nonUV\_H0Fw\_gaugeHop}); every sign pattern on the triangle (\lean{triangle\_common\_sign}), on a tree and on a ring with an odd number of sites (not formalized) is of this kind. On a bipartite graph the condition holds only for such hopping (\lean{common\_sign\_of\_shortestPathSigned\_bipartite}; Theorem 5.1 of Ref.~\onlinecite{HameedEtAl2021}), so on bipartite lattices it gives nothing beyond these sign changes; examples are chains, even rings, and square or cubic lattices with open boundaries or with periodic boundaries of even length. On other graphs it can hold for more hopping: on a complete graph it holds for every nonzero hopping (\lean{shortestPathSigned\_top}), as on every graph with unique shortest paths~\cite{HameedEtAl2021} (not formalized), and on the complete graph with four sites, hopping with $t_{12}<0$ and $t_{ij}>0$ on the other edges is not of this kind (\lean{k4\_shortestPathSigned\_not\_gauge}). The same holds for the hopping of a square lattice with periodic boundaries of odd length in both directions, with one sign along $x$ and the other sign along $y$ (not formalized): all shortest paths between two sites make the same numbers of steps along $x$ and along $y$, so the condition holds, while the two loops around the lattice, both of odd length, have hopping products of opposite signs, which no sign changes of the orbitals can make equal. The condition fails on the square $1$--$2$--$3$--$4$ with $t_{12}<0$ and $t_{23},t_{34},t_{41}>0$, where the shortest paths $1\to2\to3$ and $1\to4\to3$ have hopping products of opposite signs (\lean{square\_not\_shortestPathSigned}). There the argument does not apply, but for this square we prove (H$^*$) directly, for $0<N<4$, every interaction and all magnitudes of the hopping. For each configuration $c$ we compute the leading coefficient of $P_c(\lambda u)$ for one direction $u$; it is a polynomial in the magnitudes $|t_{ij}|$ whose terms all have the same sign, so it does not vanish, and Lean checks each of these exact certificates (\lean{hstar\_frustrated\_square}, \lean{krylovPoly\_ne\_zero\_of\_table}). The conclusions of Theorems~\ref{thm:ghk} and~\ref{thm:dens} therefore hold on this square (\lean{generic\_hk\_frustrated\_square}, \lean{dimH\_nonUVDens\_frustrated\_square}). Sign changes of the orbitals bring every other sign pattern on the square either to a common sign or to this one, so these conclusions hold on the square for every nonzero hopping (\lean{generic\_hk\_square\_of\_ne\_zero}, \lean{dimH\_nonUVDens\_square\_of\_ne\_zero}). For other graphs on which the condition fails, a general proof is open.
\end{remark}

\subsection{Non-uniquely v-representable densities}
\label{sec:densities}

Theorem~\ref{thm:ghk} is a statement about potentials. In density space, Penz and van~Leeuwen proved that the
densities produced by potentials that differ by more than a constant lie in a finite union of linear subspaces of dimension less than $M$, and hence have measure zero [Theorem 11 of
Ref.~\onlinecite{PenzvanLeeuwen2021}]. The generic theorem allows us to sharpen this result for non-degenerate ground states.\ The set of Penz and van~Leeuwen also contains densities of degenerate ground states and of ensembles. Contrary to Corollary~10 of Ref.~\onlinecite{PenzvanLeeuwen2023} (cited as a known result in Sec.~V of Ref.~\onlinecite{PenzvanLeeuwen2024}), it is not closed: for two fermions on the square, the densities $\rho_A$ of Eq.~(52) of Ref.~\onlinecite{PenzvanLeeuwen2021}, with $s$ their parameter, belong to it for $s\neq0$ and tend to $(\tfrac12,\tfrac12,\tfrac12,\tfrac12)$ as $s\to0$; we show that this density is produced only by constant potentials, even as an ensemble density (by the weak Hohenberg-Kohn argument; the Hamiltonian is real, so complex states give no other ensemble densities: the real and imaginary parts of a complex ground state, when nonzero, are real ground states). The same example is a counterexample to the conjecture of Ref.~\onlinecite{PenzvanLeeuwen2021} (their Secs.~IV\,A and VII) that the uniquely $v$-representable densities form an open subset of the density domain: the densities produced by a single potential (up to a constant) do not form such a set. This is consistent with Theorem~13 of Ref.~\onlinecite{PenzvanLeeuwen2021}, which proves openness for the uniquely $v$-representable densities of non-degenerate ground states: the limit density belongs to the degenerate ground level of the uniform square. We call a
density non-uniquely v-representable (non-uv) if it is the density of a normalized, non-degenerate ground state
$\Psi$ of $\Hop(v)$ that is also a ground state of $\Hop(v')$, with $v'-v$ not constant. The densities of $N$
particles lie in the hyperplane $\sum_i\rho_i=N$, which has dimension $M-1$.

\begin{theorem}[Non-uv densities]
\label{thm:dens}
Under the assumptions of Theorem~\ref{thm:ghk}, the set of non-uv densities has Hausdorff dimension at most
$M-3$.
\end{theorem}

The non-uv densities therefore have codimension at least two in the space of densities. Note that a set of
measure zero can have codimension one, like a surface in three dimensions. The bound is attained: for two
fermions on the square ($M=4$), the densities $\rho_A$ of Eq.~(52) of Ref.~\onlinecite{PenzvanLeeuwen2021} are
non-uv for $s\neq0$ and fill an open segment without its midpoint $(\tfrac12,\tfrac12,\tfrac12,\tfrac12)$; this set has dimension one.

The proof uses the invisible directions of Sec.~\ref{sec:setting}. Let $\Psi$ be a non-degenerate ground state
of $\Hop(v)$ with a non-uv density, and let $S$ be the set of configurations on which $\Psi$ does not vanish.
Every potential $w$ such that $\sum_i w_i n_i(c)$ is constant on $S$ leaves $\Psi$ an eigenstate of
$\Hop(v+w)$. Because of the gap, $\Psi$ remains the ground state for small $w$. These directions form a linear
space $K_S$ that contains the constants and $v'-v$, so its dimension is at least two. Moving the potential along
$K_S$ does not change the density. We can therefore move every non-uv potential close to $v$ to an affine slice
$a+W$, where $W$ is a complement of $K_S$ with dimension at most $M-2$. The moved potentials are still
exceptional, so they are zeros of $\prod_c P_c$. The point $a$ can be chosen such that this polynomial does not
vanish identically on the slice. Its zero set in the slice then has Hausdorff dimension at most $M-3$. This
follows by induction on the degree of the polynomial. Near a zero where the gradient does not vanish, the
projection of the zero set onto a hyperplane does not shrink distances by more than a fixed factor. Finally,
near a ground state with a gap the density is a Lipschitz function of the potential, and Lipschitz maps do not
increase the Hausdorff dimension. The non-degeneracy of $\Psi$ enters only through the gap. We do not know
whether the bound also holds for degenerate ground states.

\section{Spin-density functional theory}
\label{sec:spin}

For electrons with spin we take $N_\uparrow$ spin-up and $N_\downarrow$ spin-down fermions, with $0<N_\sigma<M$.
The hopping $\hat T_\uparrow+\hat T_\downarrow$ conserves spin and has the same amplitudes for both spins. The
interaction $\hat W$ is any operator diagonal in the configurations $(c_\uparrow,c_\downarrow)$, for example the
Hubbard term
$U\sum_i\hat n_{i\uparrow}\hat n_{i\downarrow}$, and spin-dependent potentials $v_{i\sigma}$. Because both
particle numbers are fixed, a constant shift of $v_\uparrow$ and a separate constant shift of $v_\downarrow$
change neither the ground states nor the spin densities. Theorem~\ref{thm:spin} assumes the setting just described.

\begin{theorem}[Generic Hohenberg-Kohn theorem for spin-DFT]
\label{thm:spin}
Assume in addition that $G$ is connected and that the hopping amplitudes have a common sign. Consider the
potentials $(v_\uparrow,v_\downarrow)$ for which some ground state is shared with a potential
$(v'_\uparrow,v'_\downarrow)$ such that $v_\uparrow-v'_\uparrow$ or $v_\downarrow-v'_\downarrow$ is not constant.
These potentials lie in a closed, nowhere dense subset of $\R^{2M}$ with zero Lebesgue measure.
\end{theorem}

The proof shows that the leading coefficient of the Krylov determinant of the two species factorizes into
the spinless leading coefficients of Sec.~\ref{sec:generic}. In the language of a charge potential $v$ and a
magnetic field $B$, with $v_\uparrow=v-B$ and $v_\downarrow=v+B$, the two constants become a constant in $v$
and a uniform shift of $B$. At fixed $N_\uparrow$ and $N_\downarrow$, a uniform shift of $B$ never changes the ground states, for any shift and any Hamiltonian. Shifting $v$ by $a$
and $B$ by $b$ adds $(a-b)\hat N_\uparrow+(a+b)\hat N_\downarrow$, which is a constant in the sector of fixed
particle numbers.
Theorem~\ref{thm:spin} says that for almost every $(v,B)$ this is the only freedom besides the constant in $v$. At fixed total particle number, as in Refs.~\onlinecite{CapelleVignale2001,EschrigPickett2001,Ullrich2005}, a uniform shift of $B$ leaves the ground state unchanged only for shifts that cause no level crossing.
The known non-uniqueness of spin-DFT~\cite{CapelleVignale2001,EschrigPickett2001} was expected by Capelle and Vignale to have a single generic source, extensive constants of motion, which in spin-DFT give the uniform shift of the field~\cite{CapelleVignale2001}. Theorem~\ref{thm:spin} confirms this on lattices at fixed $N_\uparrow$ and $N_\downarrow$, with $0<N_\sigma<M$, which excludes complete polarization, under its assumptions on the hopping and the interaction. On lattices, Ullrich found two new classes of accidental non-uniqueness for noncollinear spins, which require constraints on the ground-state spin density~\cite{Ullrich2005}. For collinear spins, Gidopoulos showed that spin potentials $(v_\uparrow,v_\downarrow)$ that differ by more than a spin-dependent constant always have different ground states, as reported in Ref.~\onlinecite{Ullrich2005}; Theorem~\ref{thm:spin} is a generic lattice form of this statement. Non-generic mechanisms exist as well. If a ground state lives only on
configurations where $\sum_i w_in_i(c)$ takes its minimum, then it remains a ground state along the half-line
$v+\varepsilon w$, $\varepsilon\ge0$. When only the total particle number is fixed, so that the polarization is free (unlike the fixed $N_\uparrow$, $N_\downarrow$ of this section), a fully polarized ground state is an example: increasing the field along its polarization does not change it~\cite{CapelleVignale2001}.

If the potential couples only to the total density $n_{i\uparrow}+n_{i\downarrow}$, the generic argument does
not apply directly, because such a potential does not break the spin symmetry of a spin-symmetric $\hat H_0$, such as the Hubbard model: states of different total spin do not mix, and for two electrons the triplet states vanish on the doubly occupied configurations for every potential, so (H$^*$) fails. We treated this case for the half-filled Hubbard dimer and for the Hubbard triangle with one spin-up and one
spin-down electron ($t=1$, $U=4$ in both cases). We split the Hamiltonian into its spin sectors and computed the Krylov
determinants of each sector in exact rational arithmetic. In both cases the total-density Hohenberg-Kohn theorem holds outside a
closed, nowhere dense set of zero measure.\ For these two examples the computation is superseded by an elementary argument, which gives more. For one spin-up and one spin-down electron with uniform hopping $t>0$, on any connected graph and for any configuration-diagonal interaction, the total-density Hohenberg-Kohn theorem holds for every potential, so the exceptional sets of both examples are empty. The configuration with the spin-up electron on site $i$ and the spin-down electron on site $j$ is the site $(i,j)$ of the product graph, with potential $v_i+v_j$ plus the interaction. No fermionic sign appears, so the single-particle theorem of Sec.~\ref{sec:examples} applies on the product graph: the ground state has no zero amplitude. The absence of fermionic signs is also the mechanism of Theorem~17 of Ref.~\onlinecite{PenzvanLeeuwen2021} for spinless fermions on the linear chain. For the Hubbard dimer the elementary argument confirms, for every $t>0$ and $U$, that the occupation difference determines the potential difference, as used by Carrascal \textit{et al.}~\cite{CarrascalEtAl2015}. The condition $t>0$ cannot be dropped: on the triangle with $t=-1$ and $U=0$, for the two potentials of Sec.~\ref{sec:examples} both electrons occupy the same one-particle ground state, so the two potentials share a non-degenerate ground state (not formalized).

\section{Kohn-Sham theory}
\label{sec:ks}

\subsection{Existence and uniqueness of the Kohn-Sham potential}

The Kohn-Sham construction needs, for a given density, a non-interacting system with that ground-state
density. On a lattice the possible densities of $N$ fermions fill the hypersimplex
$\{\rho: 0\le\rho_i\le1,\ \sum_i\rho_i=N\}$. A ground-state ensemble of $\Hop(v)$ is a mixture $\Gamma=\sum_kw_k|\Psi_k\rangle\langle\Psi_k|$ of normalized ground states $\Psi_k$, with $w_k\ge0$ and $\sum_kw_k=1$; its density is the same mixture of the densities of the $\Psi_k$.

\begin{theorem}[Existence of the Kohn-Sham potential]
\label{thm:ksexist}
For any graph and uniform hopping $t$, every density with $0<\rho_i<1$ and $\sum_i\rho_i=N$ is the density of a
ground-state ensemble of the non-interacting Hamiltonian $\hat T+\sum_i v_{s,i}\hat n_i$ for some potential
$v_s$.
\end{theorem}

The proof has two parts. First, every interior point of the hypersimplex is a strictly positive mixture of the
configuration densities $n(c)$, which we prove by induction on the number of fractional sites. Second, for any
symmetric $\hat H_0$ with a fixed particle number, a strictly positive mixture of configuration densities is
ensemble $v$-representable. The function $v\mapsto E(v)-\sum_iv_i\rho_i$ is coercive modulo constants and
attains its maximum, and a separation argument shows that at the maximum $\rho$ is the density of a
ground-state ensemble. The second part applies to interacting systems as well. The result is due to Chayes, Chayes and
Ruskai, who proved it for hypercubic lattices~\cite{ChayesChayesRuskai1985}. Penz and van~Leeuwen adapted it to
general graphs [Theorem 19 of Ref.~\onlinecite{PenzvanLeeuwen2021}]. Our contribution is a formal proof.

The Kohn-Sham potential is unique for almost every potential. Theorem~\ref{thm:ghk} with $\hat W=0$ says that,
outside a closed, nowhere dense set of zero measure, a non-interacting ground state determines the potential
up to a constant.

\begin{theorem}[Generic uniqueness of the Kohn-Sham potential]
\label{thm:ksunique}
Let the graph be connected and $t_{ij}=t_{ji}$ with $s\,t_{ij}>0$ on every edge for a fixed sign $s$. Consider the potentials $v_s$ for which a
non-interacting ground state of $\hat T+\sum_iv_{s,i}\hat n_i$ is shared with a potential differing by more than
a constant. These potentials lie in a closed, nowhere dense set of zero Lebesgue measure.
\end{theorem}

For non-interacting fermions we also prove the aufbau principle. The Slater determinant of the $N$ lowest orbitals is a ground state, with energy equal to the sum of the $N$
lowest orbital energies. A set of $N$ orbitals minimizes the sum of the orbital energies exactly when every
occupied orbital lies at or below every empty one. The Lieb functional~\cite{Lieb1983} $F(\rho)=\sup_v[E(v)-\sum_iv_i\rho_i]$ is convex on the density domain, satisfies
$E(v)\le F(\rho)+\sum_iv_i\rho_i$, with equality at the density of every ground state of $\Hop(v)$.

\subsection{Regularized Kohn-Sham iteration}

Penz \textit{et al.}~\cite{Penz2019} proposed a Kohn-Sham iteration for the Moreau-Yosida regularized Lieb
functional $F_\varepsilon(x)=\inf_\rho\big[F(\rho)+\|x-\rho\|^2/2\varepsilon\big]$, which is differentiable. The Moreau-Yosida regularization was introduced into density functional theory by Kvaal \textit{et al.}~\cite{Kvaal2014}.
Starting from a quasi-density $x_i$, the Kohn-Sham potential is
$v_{\mathrm{KS},i}=v+\nabla F_\varepsilon(x_i)-\nabla T_\varepsilon(x_i)$, where $T_\varepsilon$ is the
regularized functional of the reference (non-interacting) system. The reference system yields $x'_i$ with
$\nabla T_\varepsilon(x'_i)=-v_{\mathrm{KS},i}$, and the next iterate is
$x_{i+1}=x_i+s_i(x'_i-x_i)$ with
$s_i=-\varepsilon\langle\nabla F_\varepsilon(x_i)+v,x'_i-x_i\rangle/\|x'_i-x_i\|^2$. This is the step of
Ref.~\onlinecite{Penz2019}, where the step length $\tau_i=s_i\|x'_i-x_i\|$ multiplies the unit vector in the
direction of $x'_i-x_i$. The convergence claim of
the original paper was corrected in an erratum~\cite{Penz2020Erratum}. The steps $\|x_{i+1}-x_i\|$ tend to
zero, but this does not make the sequence convergent. We prove the corrected statement.

\begin{theorem}[Regularized Kohn-Sham iteration]
\label{thm:myksoda}
Let the interacting and the reference Hamiltonian be defined on the same lattice, and $\varepsilon>0$. Then the following holds.
(i) The regularized energies $F_\varepsilon(x_i)+\langle v,x_i\rangle$ do not increase.
(ii) The sequence $x_i$ has an accumulation point.
(iii) Every accumulation point $z$ minimizes $F_\varepsilon+\langle v,\cdot\rangle$, and the energies converge
to this minimum.
(iv) The density $\rho=z+\varepsilon v$ lies in the density domain and minimizes the Lieb energy,
$F(\rho)+\sum_iv_i\rho_i=E(v)$.
(v) Along the subsequence that converges to $z$, the Kohn-Sham potentials converge to a potential
$v_{\mathrm{KS}}$, and $\rho_{\mathrm{KS}}=z+\varepsilon v_{\mathrm{KS}}$ minimizes the Lieb energy of the
reference system in $v_{\mathrm{KS}}$.
\end{theorem}

The abstract part of the proof needs no derivatives, only two inequalities satisfied by the gradient of a
Moreau-Yosida envelope. We prove these inequalities for the lattice Lieb functional, which is convex and lower
semicontinuous on a compact, convex domain. Since $F$ is a supremum over potentials, the equality in (iv) says that $v$ maximizes $E(v')-\sum_iv'_i\rho_i$ over all $v'$. The separation argument in the proof of Theorem~\ref{thm:ksexist} then shows that $\rho$ is the density of a ground-state ensemble of $\Hop(v)$; this step is also stated in the proof of Theorem~19 of Ref.~\onlinecite{PenzvanLeeuwen2021}, where it follows from the ensemble constrained search. In the same way $\rho_{\mathrm{KS}}$ is the density of a ground-state ensemble of the reference system in $v_{\mathrm{KS}}$.

\section{Time-dependent DFT}
\label{sec:td}

\subsection{Force balance and the set $\Omega$}
\label{sec:omega}

We write the time-dependent state as $\psi=x+iy$ with real $x$ and $y$. The Schr\"odinger equation
$i\dot\psi=\Hop(v(t))\psi$ then becomes $\dot x=\Hop y$, $\dot y=-\Hop x$. The density
$\rho_i(t)=\langle\psi|\hat n_i|\psi\rangle$ obeys the continuity equation $\dot\rho_i=j_i$ with
$j_i=i\langle\psi|[\Hop_0,\hat n_i]|\psi\rangle$, where $\Hop_0=\hat T+\hat W$, and the force-balance
equation~\cite{FarzanehpourTokatly2012}
\begin{equation}
  \ddot\rho_i = q_i(\psi) - \sum_j K_{ij}(\psi)\,v_j ,
  \label{eq:forcebalance}
\end{equation}
with $q_i=-\langle\psi|[\Hop_0,[\Hop_0,\hat n_i]]|\psi\rangle$ and
$K_{ij}=\langle\psi|[\hat n_j,[\Hop_0,\hat n_i]]|\psi\rangle$. We prove both equations along every solution,
for any potential. Farzanehpour and Tokatly write the same equation as $\hat K V=\ddot\rho-q$, with
$\hat K=-K$ and the same $q$ [Eqs.~(16), (20), and (21) of Ref.~\onlinecite{FarzanehpourTokatly2012}]. We use the opposite sign, so that $K$ is positive
semidefinite at a ground state, as in Ref.~\onlinecite{CancesEtAl2026Lattice} (up to a factor 2). With a fixed particle number, $K$ annihilates the constant potentials and its columns sum to
zero. Fixing the gauge $\sum_iv_i=0$, Eq.~\eqref{eq:forcebalance} determines the potential from $\ddot\rho$
whenever the state lies in
\begin{equation}
  \Omega=\big\{\psi\ \big|\ \det\big(K(\psi)+\mathbf 1\mathbf 1^{T}\big)\neq0\big\} .
\end{equation}
Since $K$ is symmetric and annihilates the constants, this is the set of states for which $K$ has no zero mode
other than the constant potential. This is the definition of Ref.~\onlinecite{FarzanehpourTokatly2012}.
Eliminating the potential gives a nonlinear Schr\"odinger equation, to which the Picard-Lindel\"of theorem
applies.

\begin{theorem}[Local existence and uniqueness]
\label{thm:tdlocal}
Let $\Hop_0$ be any real symmetric Hamiltonian with a fixed particle number. Let $\rho(t)$ be twice
differentiable, with continuous $\ddot\rho$ and $\sum_i\ddot\rho_i=0$. Let $\psi_0\in\Omega$ be an initial state
with the density $\rho(t_0)$ and the current $\dot\rho(t_0)$. Then there is $\epsilon>0$ such that on
$(t_0-\epsilon,t_0+\epsilon)$ the density $\rho(t)$ is produced by a solution starting from $\psi_0$ and a
potential with $\sum_iv_i(t)=0$. Any two potentials with $\sum_iv_i(t)=0$ that produce $\rho(t)$ from $\psi_0$
coincide near $t_0$.
\end{theorem}

This is part (i) of Theorem 1 of Ref.~\onlinecite{FarzanehpourTokatly2012}, with the constant fixed by the
gauge. In the uniqueness statement the competing
potentials are arbitrary functions of time, not necessarily continuous.  Here and below, a potential produces $\rho(t)$ from $\psi_0$ if the Schr\"odinger equation with this potential has a solution that starts from $\psi_0$, is differentiable at every time, and has the density $\rho(t)$. Solutions that satisfy the equation only almost everywhere, which a discontinuous potential may require, are not covered. The gauge $\sum_iv_i=0$ replaces the
usual ``up to a time-dependent constant''. Removing it requires the time-dependent phase transformation of the
wavefunction, which we did not formalize.

\subsection{A counterexample for ground states}
\label{sec:counter}

For a real state $\psi=x$ and a potential $w$, let $d(c)=\sum_iw_in_i(c)$. The quadratic form of $K$ is
\begin{equation}
  \sum_{ij}w_iK_{ij}w_j = -\sum_{c,c'}(H_0)_{cc'}\,x_c\,x_{c'}\,\big[d(c)-d(c')\big]^2 ,
\end{equation}
in which the diagonal of $\Hop_0$, and therefore the interaction, drops out. At an eigenstate this equals
$2\big(\langle d\psi|\Hop|d\psi\rangle-E\langle d\psi|d\psi\rangle\big)$, an f-sum rule [Sec.~V of Ref.~\onlinecite{FarzanehpourTokatly2012}], so that at a ground
state $K$ is positive semidefinite. If $d$ is constant on the support of the ground state, then $Kw=0$. In
particular, if two potentials $v$ and $v'$ share a ground state, Eq.~\eqref{eq:support} gives $K(v-v')=0$.

Farzanehpour and Tokatly state that every ground state of a connected lattice lies in
$\Omega$~\cite{FarzanehpourTokatly2012}. This does not hold in general.

\begin{proposition}[Counterexample]
\label{prop:square}
The ground state of two fermions on the square in the potential $(\tfrac34,-\tfrac34,-\tfrac34,\tfrac34)$
(Fig.~\ref{fig:square}) has a non-constant zero mode of $K$, namely $w\propto(1,1,-1,-1)$, and therefore
does not belong to $\Omega$.
\end{proposition}

The proof of Theorem 2 in Ref.~\onlinecite{FarzanehpourTokatly2012} reduces the question to the equation
$\sum_iv_i\hat n_i\Psi_0=\lambda\Psi_0$. It then concludes that this equation ``can be fulfilled only if the function $v(r)$ takes
the same value $\lambda/N$ on all lattice sites''. The only exception mentioned there is a geometrically disconnected lattice. The
conclusion holds when $\Psi_0$ has no zero amplitude. For the state of Fig.~\ref{fig:square} the equation holds
with $v\propto(1,1,-1,-1)$ and $\lambda=0$, because both configurations on which $\sum_iv_in_i(c)\neq0$ carry
zero amplitude. The reduction to this equation, from Eq.~(61) to Eq.~(64) of Ref.~\onlinecite{FarzanehpourTokatly2012}, also assumes a non-degenerate ground state, since the terms of Eq.~(61) that belong to states degenerate with the ground state vanish. The ground state of Fig.~\ref{fig:square} is non-degenerate, so the counterexample does not depend on this. Penz and van~Leeuwen already noted that the direction $(1,1,-1,-1)$ leaves the ground state
unchanged~\cite{PenzvanLeeuwen2021}, but they did not discuss the consequence for TDDFT. In Sec.~III of Ref.~\onlinecite{FarzanehpourTokatly2012}, Farzanehpour and Tokatly noted that $K$ has additional zero modes, constant on each part, when the bonds with nonzero $t_{ij}\,\mathrm{Re}\langle\hat c_i^\dagger\hat c_j\rangle$ split a connected lattice into disconnected parts, and that an excited one-particle state of the square proposed by Kurth and Stefanucci~\cite{KurthStefanucci2011} is of this kind. The ground state of Fig.~\ref{fig:square} is of this kind as well: its bond order $\mathrm{Re}\langle\hat c_i^\dagger\hat c_j\rangle$ vanishes on the bonds 1--4 and 2--3, which splits the square into $\{1,2\}$ and $\{3,4\}$, and $w$ is the corresponding zero mode. Brown, Yang and Whitfield observed numerically that a one-electron eigenstate without full support gives
$K$ an additional zero mode, and noted that this is consistent with Ref.~\onlinecite{FarzanehpourTokatly2012}
\cite{BrownYangWhitfield2020}. The step of Theorem 2 of Ref.~\onlinecite{FarzanehpourTokatly2012} that fails also appears in the proof of the corresponding
theorem for lattice electrons coupled to cavity photons [Theorem 3 of Ref.~\onlinecite{FarzanehpourTokatly2014}]. The
relation between the two failures is direct. Every ground state shared by two potentials that differ by more
than a constant lies outside $\Omega$.

Not only the proof but also the conclusion of Theorem 2 of Ref.~\onlinecite{FarzanehpourTokatly2012} fails.

\begin{proposition}[Theorem 2 of Ref.~\onlinecite{FarzanehpourTokatly2012} does not hold]
\label{prop:ftfalse}
Let $\Psi_0$ be the ground state of Proposition~\ref{prop:square}, $w=(1,1,-1,-1)$, and
$\rho(t)=\rho(\Psi_0)+\tfrac12(t-t_0)^2\,[q(\Psi_0)-w]$. This density is smooth, has the density and the current
of $\Psi_0$ at $t_0$, and satisfies $\sum_i\ddot\rho_i=0$. No potential, continuous or not, produces it from
$\Psi_0$ in any neighborhood of $t_0$.
\end{proposition}

The proof is short. Any potential that produces the density satisfies Eq.~\eqref{eq:forcebalance} at $t_0$, so
$q(\Psi_0)-w=q(\Psi_0)-K(\Psi_0)v(t_0)$. Since $K$ is symmetric and $Kw=0$, the scalar product with $w$ gives
$\langle w,w\rangle=\langle Kw,v(t_0)\rangle=0$, which is false.

The corrected statement follows from (H$^*$).

\begin{theorem}[Ground states are generically in $\Omega$]
\label{thm:gsomega}
Consider spinless fermions on a connected graph with hopping of a common sign and any configuration-diagonal
interaction. There is a closed, nowhere dense set of static potentials of zero Lebesgue measure, outside of which
every real ground state belongs to $\Omega$.
\end{theorem}

Outside this set the TDDFT problem started from the ground state is well posed in the sense of
Theorem~\ref{thm:tdlocal}.

\subsection{Continuation in time}

The norm of the solution is conserved, so it cannot blow up. The only way the solution of
Theorem~\ref{thm:tdlocal} can stop is by approaching the boundary of $\Omega$, as stated in part (ii) of Theorem 1
of Ref.~\onlinecite{FarzanehpourTokatly2012}. Since the norm is conserved, the states lie on a compact sphere, on which a small $|\det(K+\mathbf 1\mathbf 1^T)|$ means a state close to the boundary of $\Omega$.

\begin{theorem}[Continuation]
\label{thm:global}
Under the assumptions of Theorem~\ref{thm:tdlocal} and for every $T\ge t_0$, either the solution and the
potential exist on $[t_0,T]$ with $\psi(t)\in\Omega$, or for every $\delta>0$ there is $s\in[t_0,T]$ and a
solution on $[t_0,s]$ with $|\det(K(\psi(s))+\mathbf 1\mathbf 1^T)|<\delta$.
\end{theorem}

Two solutions of the reduced (potential-eliminated) equation that stay in $\Omega$ coincide on the whole
interval. Uniqueness among arbitrary potentials, however, we have proved only locally
(Theorem~\ref{thm:tdlocal}). Whether a solution can be continued through the boundary of $\Omega$ was left open
in Ref.~\onlinecite{FarzanehpourTokatly2012}. For the singlet Hubbard dimer, Canc\`es and Duez showed that the Kohn-Sham potential
that reproduces an interacting density can be continued through the boundary of $\Omega$, but not
uniquely~\cite{CancesDuez2026Dimer}. At each such time the continuation has a free sign.

\subsection{Spin-resolved TDDFT and frozen densities}
\label{sec:spintd}

This subsection treats spinful electrons, first with the spin-resolved densities and then with the total density. With two conserved particle numbers the force-balance matrix of the spin densities $(\rho_\uparrow,\rho_\downarrow)$
always has two zero modes. We therefore work with $2M-1$ reduced observables with a single conserved sum, to
which the spinless results apply. Under the assumptions of Theorem~\ref{thm:spin}, for almost every spin-dependent static potential every real ground state is in the corresponding $\Omega$. The
spin densities determine $(v_\uparrow,v_\downarrow)(t)$ locally, up to one constant per spin (fixed by a gauge
condition).

For the total density of spinful electrons the theory can fail outright. For the total density the potential is the same for both spins, $v_\uparrow=v_\downarrow=v$. It couples to the total occupation $\hat n_{i\uparrow}+\hat n_{i\downarrow}$ of each site, and the density and the set $\Omega$ of Sec.~\ref{sec:omega} are built from these total occupations.

\begin{proposition}[Frozen densities]
\label{prop:frozen}
If an eigenstate of $\Hop_0$ has all its configurations with the same occupations, then it is an eigenstate of
$\Hop(v)$ for every $v$. Under any continuous time-dependent potential its density never changes, and, on a lattice with at least two sites, the state does not
belong to $\Omega$. For the half-filled Hubbard dimer the $S_z=0$ triplet is such a state, for every $t$ and $U$.
\end{proposition}

Farzanehpour and Tokatly remark that the dynamics of the triplet on two sites is
trivial~\cite{FarzanehpourTokatly2012}, and the spin structure of the state is fixed by the initial
state~\cite{FuksEtAl2013}. The triplet has one electron on each site in each of its configurations. A potential only multiplies it by a
phase, so the density $(1,1)$ is reproduced by every\ continuous potential. Theorem~\ref{thm:gsomega} is stated for spinless fermions, and for $t\neq0$ the triplet is not a ground state of the dimer, for any potential and any $U$, so this example does not contradict it.
However, it shows that total-density TDDFT for spinful electrons has states for which the potential is
completely undetermined.

\subsection{van~Leeuwen's theorem and time-dependent Kohn-Sham}
\label{sec:tdks}

van~Leeuwen showed that the density of an interacting system can be reproduced by a system with a different
interaction, in particular a non-interacting one~\cite{vanLeeuwen1999}. His proof assumes potentials that are analytic in time. On the lattice we prove the following, for
continuous potentials. Take any trajectory of any system, driven by a continuous potential. Its
density is reproduced locally in time by any other system on the same sites, starting from a state in its own
$\Omega$ with the same initial density and current. The potential of the second system is unique, locally in time and in the gauge $\sum_iv_i(t)=0$, among all potentials, continuous or not. This is Theorem~\ref{thm:tdlocal} applied to the second system with the density of the first. For a non-interacting second system it is noted in Ref.~\onlinecite{CancesDuez2026Dimer}, which attributes it to Refs.~\onlinecite{FarzanehpourTokatly2012,Tokatly2011}. What Theorem~\ref{thm:tdks} below adds is the set $Z$: it comes from Theorem~\ref{thm:gsomega} for the non-interacting system, and outside it every real ground state of the non-interacting system lies in $\Omega$.

\begin{theorem}[Time-dependent Kohn-Sham]
\label{thm:tdks}
For a connected graph with hopping of a common sign there is a closed, nowhere dense set $Z$ of zero measure
with the following property. Let an interacting system start from a real state, for instance its ground state,
and evolve under a continuous potential. Assume that its initial density is the density of a real non-interacting ground state of
$\hat T+\sum_iv_{s,i}\hat n_i$ with $v_s\notin Z$. Then a time-dependent Kohn-Sham potential reproduces the
interacting density near $t_0$. This potential is unique in the gauge $\sum_iv_{s,i}(t)=0$.
\end{theorem}

Real initial states carry no current, so the matching of the initial currents is automatic. The theorem also holds for a ground-state ensemble $\Gamma$ of the non-interacting system with $v_s\notin Z$: written as $\Gamma=\sum_kb_kb_k^T$, its members evolve under one common Kohn-Sham potential, unique for this decomposition. The proof purifies the ensemble into a single state on an enlarged configuration space and applies the lattice version of van~Leeuwen's theorem. This adds no case: for the pure and the ensemble statement alike, $Z$ can be chosen as the zero set of $\prod_cP_c$ for the non-interacting Hamiltonian (Sec.~\ref{sec:generic}). Outside it the spectrum is simple, so the only ground-state ensemble is the pure state $\Gamma=\psi_0\psi_0^T$. Every decomposition then has $b_k=\alpha_k\psi_0$ with $\sum_k\alpha_k^2=1$, the members evolve as $\alpha_k$ times the pure Kohn-Sham state, and the Kohn-Sham potential does not depend on the decomposition. A density that requires a mixed ensemble, as Theorem~\ref{thm:ksexist} may provide, has a degenerate non-interacting ground state, so its Kohn-Sham potential lies in $Z$, and the generic statements, Theorem~\ref{thm:tdks} and its ensemble extension, do not cover it. The set $Z$ is not empty: for two spinless fermions on the square with uniform hopping the ground level at $v_s=0$ is doubly degenerate. Farzanehpour and Tokatly anticipated an ensemble extension of their Theorems~1 and~2, for thermal states (Sec.~V of Ref.~\onlinecite{FarzanehpourTokatly2012}).

\subsection{The dimer}

\begin{figure}[t]
  \centering
  \input{figures/dimer.tex}
  \caption{One particle on two sites. All states satisfy $\dot\rho_1^2\le4T^2\rho_1\rho_2$ (shaded disk).
  The boundary is exactly the complement of $\Omega$ (bond order $b=0$). Eigenstates have zero current and lie
  inside, while the Rabi oscillation of a particle released from one site moves on the boundary.}
  \label{fig:dimer}
\end{figure}
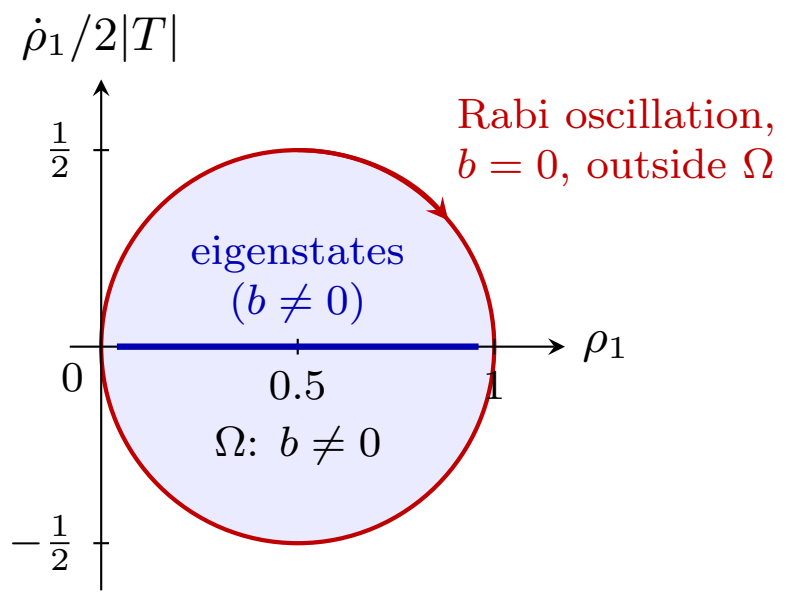

For one particle on two sites with hopping $T$ every quantity can be written explicitly, which provides a check
of the general machinery. The force-balance matrix is $K=2Tb\begin{pmatrix}1&-1\\-1&1\end{pmatrix}$ with the bond
order $b=\mathrm{Re}(\psi_1^*\psi_2)$, so that $\Omega=\{b\neq0\}$. Every state satisfies the speed limit
$\dot\rho_1^2\le4T^2\rho_1\rho_2$~\cite{LiUllrich2008,FarzanehpourTokatly2012}, with equality exactly outside
$\Omega$, as noted in Ref.~\onlinecite{FarzanehpourTokatly2012} (Fig.~\ref{fig:dimer}). We verify both
statements formally. The breakdown of the inversion for a two-site system was also discussed by
Baer~\cite{Baer2008}. The
potential that produces a given density is
\begin{equation}
  (v_1-v_2)^2\big(4T^2\rho_1\rho_2-\dot\rho_1^2\big)=\big(\ddot\rho_1+2T^2(\rho_1-\rho_2)\big)^2 .
\end{equation}
The same expression was given by Li and Ullrich [Eq.~(18) of Ref.~\onlinecite{LiUllrich2008}] and by
Farzanehpour and Tokatly [Eqs.~(47) and (50) of Ref.~\onlinecite{FarzanehpourTokatly2012}]. Every
eigenstate lies in $\Omega$, while the Rabi oscillation of a particle released from one site runs along
the boundary and never enters it.

\section{Assumptions and counterexamples}
\label{sec:assumptions}

Table~\ref{tab:assumptions} lists the main assumptions of this work. For each one we give the place where it is used
and what is known when it is dropped. Several assumptions are only technical conveniences. Others mark the
boundary of the theory, and dropping them leads to explicit counterexamples, from the literature or from this
work.

Most counterexamples have one origin, a zero amplitude of the ground state in the configuration basis. In the
continuum this cannot happen on a set of positive measure, and the unique-continuation property plays this role
in the proof of Hohenberg and Kohn~\cite{Garrigue2018}. On a lattice the analogous property can fail. The ground
state of two fermions on the triangle never occupies one pair of sites, which makes the third site full and its
potential invisible [Eqs.~(42) and (43) of Ref.~\onlinecite{PenzvanLeeuwen2021}]. On the square the ground state vanishes on
two configurations although no site is full or empty (Fig.~\ref{fig:square}). This is the state $\Psi_A$ of Eq.~(48) of Ref.~\onlinecite{PenzvanLeeuwen2021}. The same zero amplitudes make the
force-balance matrix singular (Proposition~\ref{prop:square}). Canc\`es \textit{et al.} assume a condition that excludes such
zeros, the linear independence of the states $\hat n_i\Psi$ [their Eq.~(62)], and note that it is not always true
in the discrete case~\cite{CancesEtAl2026Lattice}.

A second origin is a symmetry that is not broken by the potential. If the potential couples only to the total
density of spinful electrons, the spin symmetry of the Hubbard model is such a symmetry: some eigenstates of fixed total spin, such as the two-electron triplets, vanish on some configurations for every potential, and the generic argument does not apply. In TDDFT this produces the frozen triplet of Proposition~\ref{prop:frozen}. Degenerate singlet
and triplet ground states also lead to oscillating exchange-correlation potentials in lattice
TDDFT~\cite{RosslerVerdozziAlmbladh2018}. In the same work a density was reported as not ensemble
$v$-representable  by a non-interacting system, but Penz and van~Leeuwen later showed that it is [Sec.~VI of
Ref.~\onlinecite{PenzvanLeeuwen2024}].

Some failures cannot be removed by any assumption on the potential. For two fermions on the square, every potential on the plane
$v_1+v_3=v_2+v_4$ that is not constant is exceptional, so the generic theorems cannot be improved to
statements about every potential. The Hohenberg-Kohn theorem holds for every potential only in special cases, such as
one particle with $t>0$ (Sec.~\ref{sec:setting} and Corollary~15 of Ref.~\onlinecite{PenzvanLeeuwen2021}), one electron of each spin with uniform $t>0$ (total density, Sec.~\ref{sec:spin}), and the linear chain [Theorem 17 of
Ref.~\onlinecite{PenzvanLeeuwen2021}].

\begin{table*}[!tp]
\caption{Assumptions of lattice (TD)DFT in this work, where they are used, and what is known without them.}
\label{tab:assumptions}
\footnotesize
\raggedcells\begin{ruledtabular}
\begin{tabular}{p{4.2cm}p{5.0cm}p{7.6cm}}
Assumption & Used in & Without it \\
\hline
\multicolumn{3}{l}{\emph{Ground state}} \\
Finite, connected lattice & Single-particle HK (Sec.~\ref{sec:examples}), total-density HK for one electron of each spin (Sec.~\ref{sec:spin}),
Theorems~\ref{thm:ghk}--\ref{thm:spin}, \ref{thm:ksunique}, \ref{thm:gsomega} and~\ref{thm:tdks} & On a disconnected lattice the potential of each component has its own
  free constant~\cite{FarzanehpourTokatly2012}. \\
Filling $0<N<M$ & Rigidity of the occupations (Sec.~\ref{sec:setting}) & For $N=0$ or $N=M$ the density is fixed and the
  potential is arbitrary. \\
Fixed particle number & Constant potentials are invisible, gauge $\sum_iv_i=0$, $K$ annihilates constants & The
  grand-canonical setting is not treated here. \\
Real symmetric hopping of a common sign & Only the route sums in the proof of (H$^*$) (Sec.~\ref{sec:generic}) &
  The conclusions of Theorems~\ref{thm:ghk} and~\ref{thm:dens} also hold for nonzero hopping whose shortest paths between any two sites have hopping products of the same sign (Remark in Sec.~\ref{sec:generic}); on a bipartite lattice this is hopping that sign changes of the orbitals bring to a common sign. On the square they hold for every nonzero hopping (Remark in Sec.~\ref{sec:generic}). For other sign patterns a general proof is open. Complex hopping is not treated here. \\
Uniform hopping & Theorem~\ref{thm:ksexist}; single-particle HK (Sec.~\ref{sec:examples}) and total-density HK for one electron of each spin (Sec.~\ref{sec:spin}), both with $t>0$ & The proof of Theorem~\ref{thm:ksexist} works for any real symmetric $\hat H_0$ with a fixed particle number (Sec.~\ref{sec:ks}). \\
Interaction diagonal in the configurations & The interaction drops out of the leading coefficient
  (Sec.~\ref{sec:generic}) & Theorems~\ref{thm:ghk}--\ref{thm:spin}, \ref{thm:ksunique}, \ref{thm:gsomega} and~\ref{thm:tdks} do not cover pair hopping or correlated hopping, and for these their status is open. Theorems~\ref{thm:myksoda}, \ref{thm:tdlocal} and~\ref{thm:global}, the weak Hohenberg-Kohn theorem, the lattice van~Leeuwen theorem and the interacting system of Theorem~\ref{thm:tdks} allow any real symmetric $\hat H_0$ with a fixed particle number, so they cover them. \\
Potential couples to the site occupations & All theorems & With spin-resolved potentials, one constant per spin is
  free~\cite{CapelleVignale2001,EschrigPickett2001} (Theorem~\ref{thm:spin}), and a ground state at the minimum of
  $\sum_iw_in_i(c)$ is shared along a half-line (Sec.~\ref{sec:spin}). With coupling to the total density of spinful
  electrons, the spin symmetry of a spin-symmetric $\hat H_0$ is not broken, and (H$^*$) fails, although the Hohenberg-Kohn theorem can still hold (Sec.~\ref{sec:spin}). \\
Same hopping for both spins & Theorem~\ref{thm:spin} & Not treated here. \\
No zero amplitude of the ground state & Strong HK, Eq.~\eqref{eq:support} & Triangle [Eqs.~(42) and (43) of
  Ref.~\onlinecite{PenzvanLeeuwen2021}] and square [Eq.~(48) of Ref.~\onlinecite{PenzvanLeeuwen2021},
  Fig.~\ref{fig:square}]. Holds for almost every potential (Theorem~\ref{thm:ghk}), and for every potential for one particle with $t>0$ and for one electron of each spin with uniform $t>0$ (Sec.~\ref{sec:spin}). \\
Generic potential & Theorems~\ref{thm:ghk}, \ref{thm:spin}, \ref{thm:ksunique}, \ref{thm:gsomega} and~\ref{thm:tdks} & The
  exceptional set is not empty in general (non-constant potentials of the plane $v_1+v_3=v_2+v_4$ of the square, and one particle on the triangle with $t=-1$, Sec.~\ref{sec:examples}). Exceptions are one particle with
  $t>0$ and the linear chain~\cite{PenzvanLeeuwen2021}, and one electron of each spin with uniform $t>0$ (total density, Sec.~\ref{sec:spin}). \\
Non-degenerate ground state & Bound on the non-uv densities (Theorem~\ref{thm:dens}), through the gap, and the segment $v+\varepsilon w$, $|\varepsilon|\le g/2$, of potentials sharing the ground state on the plane of the square (Sec.~\ref{sec:examples}) & It is not known whether the bound holds for degenerate ground states (Sec.~\ref{sec:densities}). \\
Ensemble ground states, $0<\rho_i<1$ & Existence of the Kohn-Sham potential (Theorem~\ref{thm:ksexist}) & Pure-state
  $v$-representability and densities with $\rho_i\in\{0,1\}$ are not covered~\cite{ChayesChayesRuskai1985,PenzvanLeeuwen2021}.\ Time-dependent Kohn-Sham from a mixed ensemble (its static Kohn-Sham potential lies in $Z$) is not covered (Sec.~\ref{sec:tdks}). \\
\TableIcontinued
\hline
\multicolumn{3}{l}{\emph{Time-dependent}} \\
Initial state in $\Omega$ & Theorems~\ref{thm:tdlocal}, \ref{thm:global}, \ref{thm:tdks} & Ground state of the square
  (Proposition~\ref{prop:square}), which is a counterexample to Theorem 2 of Ref.~\onlinecite{FarzanehpourTokatly2012}
  (Proposition~\ref{prop:ftfalse}). Rabi state
  of the dimer (Fig.~\ref{fig:dimer}). Frozen triplet of the Hubbard dimer (Proposition~\ref{prop:frozen}). \\
Density twice differentiable, $\ddot\rho$ continuous, matching initial density and current, $\sum_i\ddot\rho_i=0$ &
  Theorem~\ref{thm:tdlocal} & These are necessary conditions. Densities above the speed limit of the dimer,
  $\dot\rho_1^2>4T^2\rho_1\rho_2$, are not produced by any
  potential~\cite{LiUllrich2008,FarzanehpourTokatly2012}. \\
Gauge $\sum_iv_i(t)=0$ & Uniqueness in Theorems~\ref{thm:tdlocal} and \ref{thm:tdks} & Replaces the arbitrary
  time-dependent constant. Removing it needs the phase transformation of the wavefunction, which is not formalized. \\
Short times & Theorem~\ref{thm:tdlocal} & The solution continues up to the boundary of $\Omega$
  (Theorem~\ref{thm:global}). For the Kohn-Sham potential of the singlet Hubbard dimer, the continuation through the
  boundary exists but is not unique~\cite{CancesDuez2026Dimer}. \\
Real initial state & Theorems~\ref{thm:gsomega} and \ref{thm:tdks} & Complex multiples follow from the invariance of
  $K$ under a global phase, which is not formalized. \\
\end{tabular}
\end{ruledtabular}
\end{table*}

\section{Formalization}
\label{sec:formal}

\subsection{What is checked}

All theorems and propositions of this paper are proved in Lean~4~\cite{Lean4} (version 4.36.0-rc2) using Mathlib~\cite{Mathlib}. The
lattice part of the code has about 17\,000 lines in 72 files, with 819 theorems and lemmas and 262 definitions.
Lean accepts a proof only if every step reduces to the axioms of its type theory. We use only the three
standard axioms (propositional extensionality, the axiom of choice, and the soundness of quotients). No step is
left unproved.

Some definitions deserve comment, because the theorems are only as good as the definitions they use. A ground
state is a nonzero minimizer of the Rayleigh quotient, written without division, which includes degenerate
ground states. The many-body Hamiltonian is a real symmetric matrix on configurations, and the potential couples
through the occupations. The fermionic hopping is the second quantization of the single-particle Hamiltonian,
defined through Slater determinants of sorted site tuples. We prove that Slater determinants of single-particle
eigenvectors are many-body eigenvectors with the sum of the orbital energies as eigenvalue, which fixes the
fermionic signs. Measure-theoretic statements use the Lebesgue measure on $\R^M$. The time evolution is the real form of the Schr\"odinger equation. The density in the force-balance equation is
the second derivative of the density of the actual solution, not an independent input. The ground states in the counterexamples are
certified by exact sum-of-squares decompositions, and the Krylov determinants of the Hubbard dimer and triangle
are evaluated in exact rational arithmetic.

\subsection{What is not checked}
\label{sec:notchecked}

The computer checks the proofs, not the correspondence between the formal statements and the physics. We read
every definition and every main statement against its physical meaning, and corrected the documentation where it
claimed more than the formal statement. The following points remain. (i) Uniqueness of the time-dependent
potential is proved in the gauge $\sum_iv_i(t)=0$ and locally in time. On a whole interval we prove uniqueness of
the solution of the reduced equation. (ii) Theorems~\ref{thm:gsomega} and~\ref{thm:tdks} are stated for real
ground states. Complex multiples follow from the invariance of $K$ under a global phase, which we did not
formalize. (iii) The non-closedness of the set of Penz and van~Leeuwen and the counterexample to the open-set conjecture of Ref.~\onlinecite{PenzvanLeeuwen2021} (Sec.~\ref{sec:densities}) are formalized, also for complex states (with ensembles taken as convex combinations of ground-state densities, as in Eq.~(2) of Ref.~\onlinecite{PenzvanLeeuwen2023}); the identification of the formal densities with $\rho_A$ of Eq.~(52) of Ref.~\onlinecite{PenzvanLeeuwen2021}, the non-degeneracy of the corresponding ground states, and the dimension of their segment are not formalized. (iv) A few side statements are marked ``not formalized'' where they appear: in the Remark of Sec.~\ref{sec:generic}, the equivalence of the formalized sign condition with its form in terms of shortest paths, and the examples of trees, odd rings, graphs with unique shortest paths and the square lattice with periodic boundaries of odd length; in Sec.~\ref{sec:spin}, the example of the triangle with $t=-1$ and $U=0$. Six further remarks are stated without a formal proof: the vanishing bond orders of the ground state of the square (Sec.~\ref{sec:counter}), that at fixed total particle number a uniform shift of $B$ leaves the ground state unchanged only for shifts that cause no level crossing (Sec.~\ref{sec:spin}), that the triplet of the Hubbard dimer is not a ground state for $t\neq0$ (Sec.~\ref{sec:spintd}), the degeneracy of the ground level of the uniform square (Secs.~\ref{sec:densities} and~\ref{sec:tdks}), that adding a constant to the potential does not change the closed set of Theorem~\ref{thm:ghk} (note after the theorem), and that a small $|\det(K+\mathbf 1\mathbf 1^T)|$ means a state close to the boundary of $\Omega$ (before Theorem~\ref{thm:global}).

\subsection{How the proofs were produced}

The Lean code was written with the assistance of an AI system (Claude, Anthropic), which also produced a first
draft of this manuscript, as a technical step to make sure that the results stated in the text match those proved in the Lean code, and assisted in its revision. The Lean kernel guarantees that every proof is correct. The choice of the problems, the mathematical strategy, and the check that the formal statements say what the text claims are the responsibility of the authors.

\section{Conclusions}
\label{sec:conclusions}

In conclusion, we proved that the failures of the Hohenberg-Kohn theorem on lattices, and of local existence and uniqueness of the time-dependent potential in lattice TDDFT for a system that starts in a real ground state, are exceptional. For spinless
fermions on any connected lattice, at any filling other than empty or full, with hopping of a common sign and any density-density
interaction, they occur only on a closed, nowhere dense set of potentials of zero measure. No non-degeneracy of the ground state is assumed here: outside this closed set every ground state is non-degenerate, and hence real up to a global phase; degeneracies, whether due to a lattice symmetry or accidental, occur only inside it. The densities of non-degenerate ground states shared by two potentials that differ by more than a constant form a set of codimension at least two. For two fermions on the square, the densities produced by two potentials that differ by more than a constant do not form a closed set, contrary to Corollary~10 of Ref.~\onlinecite{PenzvanLeeuwen2023}; the same example is a counterexample to the open-set conjecture of Ref.~\onlinecite{PenzvanLeeuwen2021}. A ground state shared by two such potentials has a zero amplitude in the configuration basis. One algebraic statement about Krylov spaces, which we proved with a strong-potential limit, shows that outside the zero set of one nonzero polynomial in the potential the spectrum is simple and no eigenstate has a zero amplitude, which excludes both failures. The same argument gives the generic Hohenberg-Kohn theorem for spin-density functional theory, where a uniform shift of the magnetic field is the only additional freedom, the corresponding statement for spin-resolved TDDFT, the generic uniqueness of the static Kohn-Sham potential, and, locally in time, the generic existence and uniqueness of the time-dependent one. Theorem~2 of Ref.~\onlinecite{FarzanehpourTokatly2012} rests on the statement that every ground state of a connected lattice belongs to the set of admissible
initial states. We found that this fails for some ground states: the ground state of two fermions on the square given by Penz and van~Leeuwen is a counterexample. For this state the conclusion of the theorem does not hold either, since a density that satisfies all its conditions is produced by no potential. All theorems and propositions are formalized in Lean~4 and checked by computer; Sec.~\ref{sec:notchecked} lists what is not.

Several questions remain open. Natural next steps are the generic theorem for hopping with two shortest paths between the same sites whose hopping products have opposite signs, such as a ring with an even number of sites, six or more, and one negative bond (the square is treated in the Remark in Sec.~\ref{sec:generic}), the bound of Theorem~\ref{thm:dens} for degenerate ground states, and the uniqueness of the time-dependent
potential on a whole interval among arbitrary potentials. For the total density of spinful electrons, the Hohenberg-Kohn theorem is proved here only for one electron of each spin with uniform hopping $t>0$. Other particle numbers are not treated here; for other hopping we give only the example of the triangle with $t=-1$ and $U=0$, where the theorem fails for some potentials (Sec.~\ref{sec:spin}).

\begin{table*}[!tp]
\caption{Main results and the names of the corresponding Lean theorems in Ref.~\onlinecite{leandft}.}
\label{tab:results}
\footnotesize \endgraf\noindent\raggedright A dagger ($^\dagger$) marks the results that, to our knowledge, are new in this work. A citation after a result indicates that the result, or a statement from which it follows directly, was given before in that reference; the contribution of this work is then the formal proof, and the text gives the relation to the earlier statement. Rows with neither mark are standard results, such as the aufbau principle and the properties of the Lieb functional, ingredients of the proofs, examples, or results whose relation to earlier work is discussed in the text.\endgraf
\raggedcells\begin{ruledtabular}
\begin{tabular}{p{6.4cm}cp{9.2cm}}
Result & Section & Lean name \\
\hline
Weak Hohenberg-Kohn theorem~\cite{PenzvanLeeuwen2021} & \ref{sec:setting} & \lean{weak\_hk} \\
Single-particle HK ($t>0$)~\cite{PenzvanLeeuwen2021}, frustrated counterexample & \ref{sec:setting} & \lean{hk\_single\_particle}, \lean{hk\_fails\_frustrated}, \lean{hk\_fails\_frustrated\_nondeg} \\
Two fermions on the triangle and on the square~\cite{PenzvanLeeuwen2021}; for the square, the non-closedness of the exceptional set and a necessary condition on the hopping\unskip$^\dagger$ & \ref{sec:setting} & \lean{hk\_fails\_triangle}, \lean{hk\_fails\_square}, \lean{square\_wall}, \lean{hk\_fails\_square\_nondeg}, \lean{square\_wall\_unit}, \lean{square\_wall\_plus}, \lean{square\_wall\_minus}, \lean{square\_plane\_exceptional}, \lean{square\_const\_not\_exceptional}, \lean{eigen\_sq\_plus}, \lean{square\_segment\_plus}, \lean{square\_example\_segment}, \lean{Bridge.Square.square\_zero\_pattern\_hopping}, \lean{Bridge.Square.square\_zero\_pattern\_hopping\textquotesingle}, \lean{Bridge.Square.square\_shared\_hopping\_sharp} \\
Generic HK, Theorem~\ref{thm:ghk} \unskip$^\dagger$ & \ref{sec:generic} & \lean{generic\_hk\_lattice\_fermions\_w}, \lean{generic\_hk\_lattice\_fermions}, \lean{generic\_hk\_lattice\_fermions\_shortest}, \lean{generic\_hk\_lattice\_fermions\_gauge}, \lean{density\_determines\_potential}, \lean{square\_exceptional\_not\_closed} \\
Sign changes of the orbitals & \ref{sec:generic} & \lean{gamma\_signConj}, \lean{H0Fw\_gaugeHop}, \lean{nonUV\_H0Fw\_gaugeHop}, \lean{triangle\_common\_sign} \\
Hopping with shortest paths of one sign~\cite{HameedEtAl2021} & \ref{sec:generic} & \lean{shortestPathSigned\_of\_gauge}, \lean{common\_sign\_of\_shortestPathSigned\_bipartite}, \lean{shortestPathSigned\_top}, \lean{k4\_shortestPathSigned\_not\_gauge}, \lean{square\_not\_shortestPathSigned} \\
Square with one negative bond; every nonzero hopping on the square$^\dagger$ & \ref{sec:generic} & \lean{hstar\_frustrated\_square}, \lean{generic\_hk\_frustrated\_square}, \lean{dimH\_nonUVDens\_frustrated\_square}, \lean{generic\_hk\_square\_of\_ne\_zero}, \lean{dimH\_nonUVDens\_square\_of\_ne\_zero}, \lean{krylovPoly\_ne\_zero\_of\_table} \\
Non-uv densities, Theorem~\ref{thm:dens} \unskip$^\dagger$ & \ref{sec:generic} & \lean{dimH\_nonUVDens\_lattice\_fermions\_w}, \lean{dimH\_nonUVDens\_lattice\_fermions}, \lean{dimH\_nonUVDens\_lattice\_fermions\_w\textquotesingle}, \lean{dimH\_nonUVDens\_lattice\_fermions\textquotesingle}, \lean{dimH\_nonUVDens\_lattice\_fermions\_shortest\textquotesingle}, \lean{dimH\_zeroSet\_add\_one\_le}, \lean{square\_sharedDens\_not\_closed}, \lean{square\_sharedEnsDens\_not\_closed}, \lean{square\_uvDens\_not\_relOpen}, \lean{square\_sharedEnsDensC\_not\_closed}, \lean{square\_uvDensC\_not\_relOpen} \\
Krylov criterion~\cite{JamingMayeli2026}, null zero sets & \ref{sec:generic} & \lean{nonUV\_imp\_krylov\_zero}, \lean{volume\_zeroSet}, \lean{generic\_hk\_ae} \\
(H$^*$) for lattice fermions & \ref{sec:generic} & \lean{hstar\_lattice\_fermions\_w}, \lean{hstar\_lattice\_fermions\_shortest}, \lean{excited\_no\_zero\_and\_simple} \\
Field shift~\cite{CapelleVignale2001,EschrigPickett2001}; spin-DFT, Theorem~\ref{thm:spin}$^\dagger$ & \ref{sec:spin} & \lean{generic\_hk\_spin}, \lean{generic\_hk\_spin\_field}, \lean{groundState\_field\_shift} \\
Hubbard model (spin-resolved); dimer, triangle and one electron of each spin (total density) & \ref{sec:spin} & \lean{generic\_hk\_hubbard}, \lean{hubbard\_dimer\_generic\_hk\_ae}, \lean{hubbard\_triangle\_generic\_hk\_ae}, \lean{Bridge.Hubbard.H0S\_dimer}, \lean{Bridge.Hubbard.H0S\_triangle}, \lean{Bridge.Hubbard.occD\_eq\_occF}, \lean{Bridge.Hubbard.occTr\_eq\_occF}, \lean{two\_electron\_hk}, \lean{Bridge.TwoElectron.hamN\_H0S\_eq\_ham2}, \lean{Bridge.TwoElectron.two\_electron\_hk\_H0S}, \lean{Bridge.TwoElectron.two\_electron\_density\_determines\_potential}, \lean{Bridge.TwoElectron.two\_electron\_ground\_ne\_zero} \\
Half-line non-uniqueness & \ref{sec:spin} & \lean{nonunique\_of\_minimiser} \\
Kohn-Sham existence, Theorem~\ref{thm:ksexist}~\cite{ChayesChayesRuskai1985,PenzvanLeeuwen2021} & \ref{sec:ks} & \lean{ks\_potential\_exists\_interior}, \lean{ensemble\_v\_representable} \\
Kohn-Sham uniqueness, Theorem~\ref{thm:ksunique} \unskip$^\dagger$ & \ref{sec:ks} & \lean{ks\_potential\_generic\_unique}, \lean{ks\_potential\_generic\_unique\_w} \\
Aufbau, Lieb functional & \ref{sec:ks} & \lean{aufbau}, \lean{aufbau\_lowest}, \lean{slaterSub\_eigen}, \lean{qf\_gamma\_ge}, \lean{lieb\_duality}, \lean{lieb\_convexOn} \\
Regularized Kohn-Sham iteration, Theorem~\ref{thm:myksoda}~\cite{Penz2019,Penz2020Erratum} & \ref{sec:ks} & \lean{lattice\_myksoda}, \lean{lattice\_myksoda\_ensemble}, \lean{lattice\_myksoda\_exists} \\
\TableIIcontinued
Continuity and force balance~\cite{FarzanehpourTokatly2012} & \ref{sec:td} & \lean{tdens\_hasDerivAt}, \lean{tcur\_hasDerivAt} \\
Local existence and uniqueness, Theorem~\ref{thm:tdlocal}~\cite{FarzanehpourTokatly2012} & \ref{sec:td} & \lean{td\_existence}, \lean{td\_uniqueness} \\
Counterexample, Propositions~\ref{prop:square} and \ref{prop:ftfalse} \unskip$^\dagger$ & \ref{sec:td} & \lean{square\_kMat\_degenerate}, \lean{kMat\_sq\_w}, \lean{kfb\_sq\_w}, \lean{square\_not\_mem\_omega}, \lean{ft\_theorem2\_counterexample}, \lean{hk\_fails\_square\_nondeg}, \lean{shared\_ground\_state\_not\_mem\_omega}, \lean{shared\_ground\_state\_pair\_not\_mem\_omega}, \lean{prop9\_paper} \\
Ground states in $\Omega$, Theorem~\ref{thm:gsomega} \unskip$^\dagger$ & \ref{sec:td} & \lean{generic\_ground\_mem\_omega} \\
Continuation, Theorem~\ref{thm:global}~\cite{FarzanehpourTokatly2012} & \ref{sec:td} & \lean{td\_global}, \lean{goodOn\_unique} \\
Spin-resolved TDDFT \unskip$^\dagger$ & \ref{sec:td} & \lean{generic\_ground\_mem\_omega\_spin}, \lean{td\_spin\_existence}, \lean{td\_spin\_uniqueness} \\
Frozen densities, Proposition~\ref{prop:frozen} & \ref{sec:td} & \lean{hamN\_mulVec\_frozen}, \lean{not\_mem\_omega\_of\_frozen}, \lean{hubbard\_triplet\_frozen}, \lean{tdens\_of\_frozen}, \lean{hubbard\_triplet\_frozen\_unique}, \lean{tdens\_of\_frozen$_2$}, \lean{not\_mem\_omega\_of\_occ\_const} \\
van~Leeuwen's theorem~\cite{FarzanehpourTokatly2012,CancesDuez2026Dimer}; TD Kohn-Sham, Theorem~\ref{thm:tdks}$^\dagger$ & \ref{sec:td} & \lean{van\_leeuwen}, \lean{van\_leeuwen\_unique}, \lean{td\_kohn\_sham}, \lean{td\_kohn\_sham\_ensemble}, \lean{Ens\_eq\_singleton\_of\_hstar}, \lean{generic\_Ens\_pure}, \lean{decomp\_pure}, \lean{td\_kohn\_sham\_ensemble\_pure} \\
Dimer~\cite{LiUllrich2008,FarzanehpourTokatly2012} & \ref{sec:td} & \lean{mem\_omega\_dimer\_iff}, \lean{speed\_limit\_eq\_iff}, \lean{dimer\_td\_potential} \\
\end{tabular}
\end{ruledtabular}
\end{table*}

\begin{acknowledgments}
This work was carried out with the assistance of the AI system Claude (Anthropic) for the development of the Lean
code and the preparation of the manuscript (see Sec.~\ref{sec:formal}).
\end{acknowledgments}

\section*{Data and Software Availability}
The Lean code, together with a human-readable blueprint of all definitions and theorems, is openly available at
\url{https://github.com/hyllios/lean_dft} and \url{https://hyllios.github.io/lean_dft/}.

\bibliography{references}

\end{document}

%% file: figures/square.tex
\begin{tikzpicture}[
    site/.style={circle, draw, thick, minimum size=7mm, inner sep=0pt, fill=white},
    conf/.style={rounded corners=2pt, draw, minimum width=9mm, minimum height=5.5mm,
                 inner sep=1pt, fill=white, font=\footnotesize},
    zero/.style={conf, draw=red!75!black, text=red!75!black, thick},
    lab/.style={font=\footnotesize}]
  \node[font=\small] at (-0.4,2.55) {(a)};
  \coordinate (s1) at (0,1.6);
  \coordinate (s2) at (1.6,1.6);
  \coordinate (s3) at (1.6,0);
  \coordinate (s4) at (0,0);
  \draw[thick] (s1) -- (s2) -- (s3) -- (s4) -- cycle;
  \node[site] at (s1) {1};
  \node[site] at (s2) {2};
  \node[site] at (s3) {3};
  \node[site] at (s4) {4};
  \node[lab, above=3.5mm] at (s1) {$\tfrac{3}{4}\,|\,\tfrac{5}{4}$};
  \node[lab, above=3.5mm] at (s2) {$-\tfrac{3}{4}\,|\,-\tfrac{1}{4}$};
  \node[lab, below=3.5mm] at (s3) {$-\tfrac{3}{4}\,|\,-\tfrac{5}{4}$};
  \node[lab, below=3.5mm] at (s4) {$\tfrac{3}{4}\,|\,\tfrac{1}{4}$};
  \node[lab, left=3.5mm] at (s1) {$\rho_1=\tfrac{1}{5}$};
  \node[lab, right=3.5mm] at (s2) {$\rho_2=\tfrac{4}{5}$};
  \node[lab, right=3.5mm] at (s3) {$\rho_3=\tfrac{4}{5}$};
  \node[lab, left=3.5mm] at (s4) {$\rho_4=\tfrac{1}{5}$};
  \begin{scope}[xshift=5.4cm, yshift=0.8cm]
    \node[font=\small] at (-2.0,1.75) {(b)};
    \node[conf] (c13) at (-1.45,0)  {13: $2$};
    \node[conf] (c24) at (1.45,0)   {24: $2$};
    \node[zero] (c12) at (0,1.5)    {12: $0$};
    \node[conf] (c14) at (0,0.5)    {14: $1$};
    \node[conf] (c23) at (0,-0.5)   {23: $4$};
    \node[zero] (c34) at (0,-1.5)   {34: $0$};
    \foreach \c in {c12,c14,c23,c34} {\draw (c13) -- (\c); \draw (c24) -- (\c);}
  \end{scope}
\end{tikzpicture}

%% file: figures/proof.tex
\begin{tikzpicture}[
    box/.style={draw, rounded corners=2pt, align=center, font=\footnotesize,
                text width=3.6cm, inner sep=3pt, fill=white},
    res/.style={box, fill=blue!7, thick},
    >=stealth, node distance=4mm]
  \node[box] (zero) {shared ground state, $v-v'$ non-constant\\$\Rightarrow$ zero amplitude $\Psi(c)=0$};
  \node[box, below=of zero] (kry) {Krylov criterion:\\$P_c(v)=\det[\,|c\rangle,\hat H|c\rangle,\dots]=0$};
  \node[box, below=of kry] (null) {zero set of a polynomial\\$P\not\equiv 0$: closed, nowhere dense,\\Lebesgue null};
  \node[box, right=6mm of zero] (top) {$v=\lambda u$: leading coefficient\\of $P_c$ independent of $\hat W$};
  \node[box, below=of top] (slater) {free fermions (Slater):\\signed sum over optimal\\assignments of route sums};
  \node[box, below=of slater] (res) {resonance: one assignment\\dominates $\Rightarrow$ $P_c\not\equiv0$ (H$^*$)};
  \node[res, below=7mm of null, xshift=2.1cm] (hk) {generic HK (Thm.~\ref{thm:ghk}), non-uv densities (Thm.~\ref{thm:dens}), spin (Thm.~\ref{thm:spin}),\\
        Kohn--Sham (Thm.~\ref{thm:ksunique}), TDDFT from the ground state (Thm.~\ref{thm:gsomega}), time-dependent Kohn--Sham (Thm.~\ref{thm:tdks})};
  \draw[->] (zero) -- (kry);
  \draw[->] (kry) -- (null);
  \draw[->] (top) -- (slater);
  \draw[->] (slater) -- (res);
  \draw[->] (res.south) |- ([yshift=2mm]hk.north east) ;
  \draw[->] (null.south) -- (null.south |- hk.north);
\end{tikzpicture}

%% file: figures/dimer.tex
\begin{tikzpicture}[scale=2.6, >=stealth]
  \fill[blue!8] (0.5,0) ellipse (0.5 and 0.5);
  \draw[thick, red!75!black] (0.5,0) ellipse (0.5 and 0.5);
  \draw[->] (-0.08,0) -- (1.18,0) node[right] {$\rho_1$};
  \draw[->] (0,-0.62) -- (0,0.68) node[above] {$\dot\rho_1/2|T|$};
  \foreach \x in {0.5,1} \draw (\x,0.02) -- (\x,-0.02) node[below, font=\footnotesize] {$\x$};
  \node[below left, font=\footnotesize] at (0,0) {$0$};
  \draw (0.02,0.5) -- (-0.02,0.5) node[left, font=\footnotesize] {$\tfrac12$};
  \draw (0.02,-0.5) -- (-0.02,-0.5) node[left, font=\footnotesize] {$-\tfrac12$};
  \draw[very thick, blue!70!black] (0.04,0) -- (0.96,0);
  \node[font=\footnotesize, blue!70!black, align=center] at (0.5,0.17) {eigenstates\\($b\neq0$)};
  \node[font=\footnotesize, align=center] at (0.5,-0.25) {$\Omega$: $b\neq 0$};
  \draw[->, thick, red!75!black] (0.5,0.5) arc[start angle=90, end angle=40, x radius=0.5, y radius=0.5];
  \node[font=\footnotesize, red!75!black, align=left, anchor=west] at (0.86,0.52)
       {Rabi oscillation,\\ $b=0$, outside $\Omega$};
\end{tikzpicture}